\documentclass{sage}
\usepackage{amsthm,amssymb,natbib,url,amsmath}
\usepackage{subfigure}

\begin{document}

\title{Dynamic Homeostasis in Packet Switching Networks}
\author{Mizuki Oka \thanks{Corresponding author;
e-mail: mizuki@cs.tsukuba.ac.jp} and Hirotake Abe}
\address{Department of Computer Science, University of Tsukuba, Japan}
\author{Takashi Ikegami}
\address{Department of General Systems Studies, The University of Tokyo, Japan}

\maketitle

\begin{abstract}
In this study, we investigate the adaptation and robustness of a packet switching network (PSN), the fundamental architecture of the Internet. We claim that the adaptation introduced by a transmission control protocol  (TCP) congestion control mechanism is interpretable as the self-organization of multiple attractors and stability to switch from one attractor to another. To discuss this argument quantitatively, we study the adaptation of the Internet by simulating a PSN using ns-2. Our hypothesis is that the robustness and fragility of the Internet can be attributed to the inherent dynamics of the PSN feedback mechanism called the congestion window size, or \textit{cwnd}. By varying the data input into the PSN system, we investigate the possible self-organization of attractors in cwnd temporal dynamics and discuss the adaptability and robustness of PSNs. The present study provides an example of Ashby's Law of Requisite Variety in action.
\end{abstract}


\section{Introduction}
What is homeostasis of the Internet?  Homeostasis is an important concept proposed by Cannon~\citep{WBCannon} as a biological mechanism for sustaining stability.  The concept has been developed and applied to many other systems, ranging from artificially designed systems to ecological and the Gaia systems. As we discuss in the following, to explore the robustness and fragility of the Internet, it is necessary to examine the idea of homeostasis in the Internet.

\subsection{Homeostasis and Ashby's Law}
In 1930, W. B. Cannon~\citep{WBCannon} introduced the idea of homeostasis as a mechanism that maintains the internal state of a system (e.g., calcium levels in the blood, body temperature, or immune responses). Simply put, such a mechanism could be explained using the negative feedback loop within a system. For example, in the case of a healthy human body, if the body temperature increases beyond a certain threshold, the body tries to lower the temperature by sweating. If calcium levels in the blood are too low, the body tries to generate calcium to maintain a certain percentage of calcium in the blood. However, in certain circumstances, it is possible to maintain system functions by intentionally deviating from the steady state. This is system-level homeostasis rather than component-level homeostasis. A few studies have constructed a theoretical background for this concept. Homeostasis is not merely an attracting state of dynamical systems; rather, it has to be adaptive. Primarily, a homeostatic system requires an environment. 
A system has some functionality (e.g., perception or motor outputs)  in the environment and homeostasis means to preserve the functionality of the system. Namely, it provides a control theory of a complex system.
Secondly, each component (either species or processes) of a homeostatic system should work in concert with the other components without being controlled by a central unit. This second criterion provides an another notion of symbiotic relationship among independent species.

The former point, homeostasis as a control theory, is developed by 
 Ashby as the Law of Requisite Variety~\citep{ashby_1958, Umpleby_2009}. This law states that an active controller requires at least as many states as exist in the controlled system in order to be stable. It attempts to demonstrate that the effective degrees of freedom of the system change adaptively in response to external perturbations in order to control a system. Ashby compared the law with Shannon's information theory (more specifically, with successful signal transmission in a noisy channel) and argued for a control theory with local and global feedback loops, which he termed the \textit{ultra-stability}~\citep{ashby1960, homeostats_franchi2013}. According to this theory, when a system reaches a critical condition (which is measured by the relevant essential variable), it changes its behavior. It suddenly randomly alters its dynamic parameters to update its global behavior until it escapes the critical condition. This random updating of parameters to restore a functional state is called \textit{ultra-stable control}. The ultra-stable control is an extended example of the Law of Requisite Variety.

Di Paolo studied biological adaptation as a new version of Ashby's ultra-stability. Using the example of a visual field inversion experiment (or, equivalently, the upside-down glasses experiment)~\citep{george_1896}, he argued that biological adaptation is different from mechanical adjustments, which do not have internal dynamics~\citep{paolo_2000}. When glasses are reversed, the vertical and horizontal visual perspectives are reversed as well. However, within a week or so, a human's perception begins to return to its normal state. As Di Paolo argued, this outstanding adaptability, called \textit{homeo-adaptation}, is facilitated by plastic synapse dynamics, in which no synaptic change occurs when the neural firing rate is in the middle range~\citep{paolo_2000}. Iizuka and Di Paolo used the same idea to show how a simple mobile agent restores a phototaxis~\citep{paolo_2007} being perturbed by sensory-motor coupling.

The latter point, homeostasis as an ecosystem, is developed by Lovelock and Margulis as the Gaia hypothesis, which considers the entire planet as a large symbiotic system~\citep{lovelock1974}. The Gaia theoretical model is constructed as the Daisyworld~\citep{daisyworld}, and the extension of this model has been studied by many researchers~\citep{homeostat_inman2004,daisyworld_review,suzuki2008,dyke2013}. Briefly, in the Daisyworld model, black and white daisies, as a net result, can regulate local temperatures by changing their population sizes. Because black and white daisies have different albedos (reflections of light), when the local temperature increases, white daisies outgrow black daisies. Conversely, as the temperature drops, black daisies outgrow white daisies. This competition leads to a regulation of the surrounding temperature. The Gaia theory has been revisited through a new modeling approach, which uses the context of evolution of a biosphere~\citep{lenton_nature,andrew_gaia}. However, the interactions within a large ecosystem and the resulting self-regulating mechanisms need to be investigated further.

The essential mechanism of homeostasis in either control theory or the theory of ecosystem involves maintaining instability to organize adaptation (for example, Ashby's randomly updated system parameters or Di Paolo's ignition of synaptic plasticity). To demonstrate this, Ikegami and Suzuki developed a mobile robot that placed daisies on the robot's surface and showed that the robot moved by itself because of surface temperature regulation~\citep{suzuki2008}. Maintaining instability helped it accommodate changes in the environment. Also the homeostatic mechanism by a large but weak chaotic state is discussed by Kaneko and Ikegami~\citep{ike90}, which is an example of symbiosis by maintaining instability. 
This mechanism of organizing inherent instability is the main theme of this paper. The Internet is an interesting case for studying such homeostatic characteristics because it establishes stable control system on top of an open-ended system with constantly changing volumes of external input; stability on top of unstable dynamics ~\citep{ecal_psn}. 

\subsection{Homeostasis in the Interenet}
The Internet system consists of a massive number of computers and signal transmission cables, such that each composite computer is a purposeful unit that receives and sends information to other computers. Yet, though the Internet changes quickly, it preserves a certain structure over time. One study has shown that nearly half of the pages on the web become unavailable after a year~\citep{Fetterly03,Ntoulas04}. At the same time, new pages are constantly created, either by people or by automated bots. These pages are crawled by search bots, indexed by search engines, and made accessible to users. If a system is too rigid to change its behavior according to changes in external inputs, no adaptation is expected. Moreover, if a system is governed only by external changes and is unable to restore its original behavior, it is still not an adaptive system; rather, it has a single, stubborn, attracting state. In contrast, the Internet is capable of constantly exchanging data by adapting to environmental changes. This "ecosystem" of the Internet has been established over the past several decades.

While the Internet exhibits a great deal of robustness, its fragility, too, has been discussed.  Robustness refers to a generic property of stability against perturbation. Fragility, despite common perception, is not an opposite property. For example, a dead state is also an example of robust state, but it is not a homeostatic state. What, then, would be the biological sense of homeostasis? For example, in order to take in environmental information, a system should have fragility built into it. Taking in environmental information means lowering entropy of some sort, and doing this requires instability. However, a system would not function if the entire system became unstable due to the incorporation of fragility. Some type of balancing mechanism is required. In fact, the robustness of the Internet is derived from its fragility~\citep{RYF_pnas}. This Robust Yet Fragile (RYF) characteristic of the Internet is often explained using the scale-free feature~\citep{RYF_pnas}. A "hub-like" core structure makes the network simultaneously robust to random losses of nodes and fragile to targeted attacks on highly connected nodes, or "hubs"~\citep{barabasi}. However, this explanation is focused on the properties of topological network structures, and it does not account for other important factors, such as feedback mechanisms or dynamic packet transmission routing. The Internet is also a collective of massive data flows and the certification of successful signal transmissions through the devising of signal transmission protocols. Namely, the protocols adaptively change channel capacity using an acknowledgement signal. This is called a \textit{packet switching network} (PSN). How can the RYF property of the PSN can be explained in terms of homeostasis?

To quantitatively study this notion of homeostasis using the Internet, we examine a PSN simulator called ns-2 and discuss its adaptability and robustness. PSNs constitute the fundamental mechanisms of the Internet, which provide adaptive dynamics to the system. We consider that the adaptation of a PSN involves multiple attractors, as well as the ability to switch among attractors. Given a set of nodes in the network, a PSN ensures that data, which are divided into \textit{packets}, are sent safely to the Internet. We argue that a PSN shows adaptation to the system (i.e., adjustments in response to packet congestion in the system) via the self-organization of many attractors. Here, adaptation means keeping the throughput high by changing the dynamics (i.e., the number of packets to be sent per unit of time from each node).  

The robustness of a PSN is based on the number of packets each node can send; this is called the congestion window size, or \textit{cwnd}. The cwnd of each node changes according to the congestion state of the network. The global congestion state of the entire system indirectly feeds back into the local cwnd dynamics of each node. The self-organization mechanism of the PSN is the underlying mechanism that enables the system to maintain a certain throughput, thus making the system efficient. By varying the amount of data input to the PSN system, we investigate the possible self-organization of attractors under cwnd temporal dynamics and discuss the PSN's adaptability. In order to discuss the dynamic characteristics of PSN, we define a PSN state that changes over time. Candidates for a PSN state at time $t$ could be a set of cwnd size, a set of buffer size, the dropping events of packets, the flight time for each packet, and so on. These cwnd sizes sometimes synchronize and desynchronize over time; they also bifurcate through changes in the input values. In the following, we first see the dynamics of PSN naively by taking the aforementioned variables as PSN states; then, we translate the variables into the principle components and take the first and second principal components as new PSN states in order to examine their attracting states and the transitions among different attractors. The results presented in this study reflect Ashby's Law of Requisite Variety in action and provide a concrete example to support the discussion.

The rest of this paper is organized as follows: Section 2 describes PSN mechanisms and the ns-2 simulator and shows the manner in which system behavior changes in response to variations in external inputs or system perturbations. Section 3 focuses on the organization of PSN states and the dynamics of the attracting states in relation to system throughput. In Section 4, the findings of this study are described and the paper is concluded.

\section{Packet Switching Networks}
PSNs constitute the fundamental architecture of the Internet, which consists of nodes and edges that form a network. Each node in the network follows an algorithm implemented in the transmission control protocol (TCP). TCP plays an important role in delivering data to their respective destinations without packet drops or permutations. Furthermore, it controls the sending and resending of lost packets. Generally, a node starts by sending packets slowly. When another node receives a sent packet, it sends back a sequence number called an acknowledgement (ACK) to confirm receipt of the packet. When the sender node receives this ACK, it sends the next packet. If the sender node does not receive an ACK within a certain amount of time or if it receives ACKs with unexpected sequence numbers, it resends the same packet.

\begin{figure}[!ht]
\centering
\includegraphics[height=0.12\textwidth]{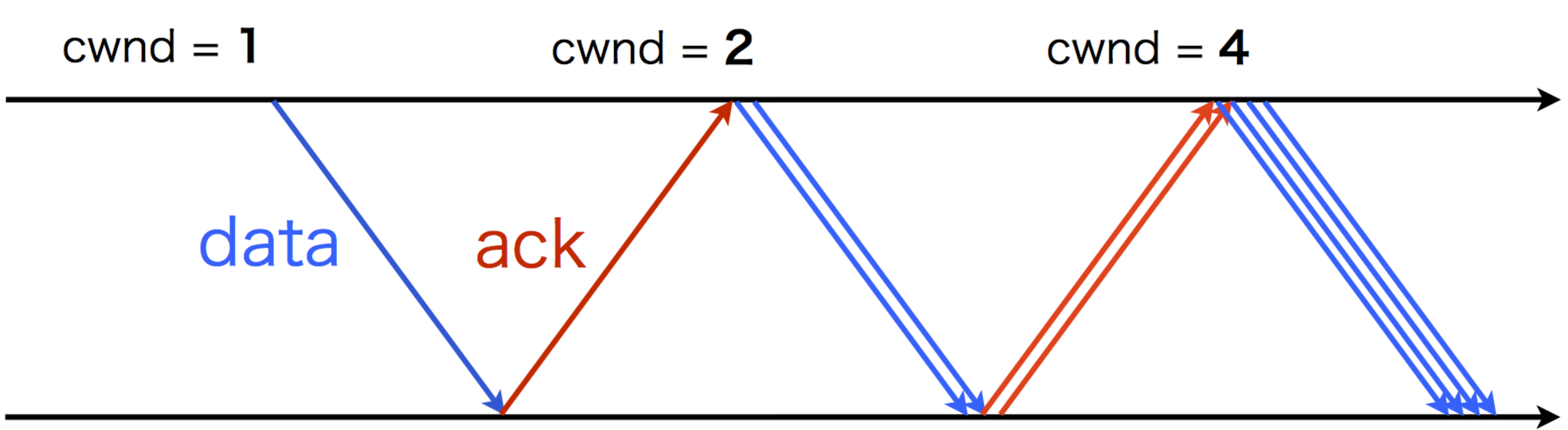}
\includegraphics[height=0.12\textwidth]{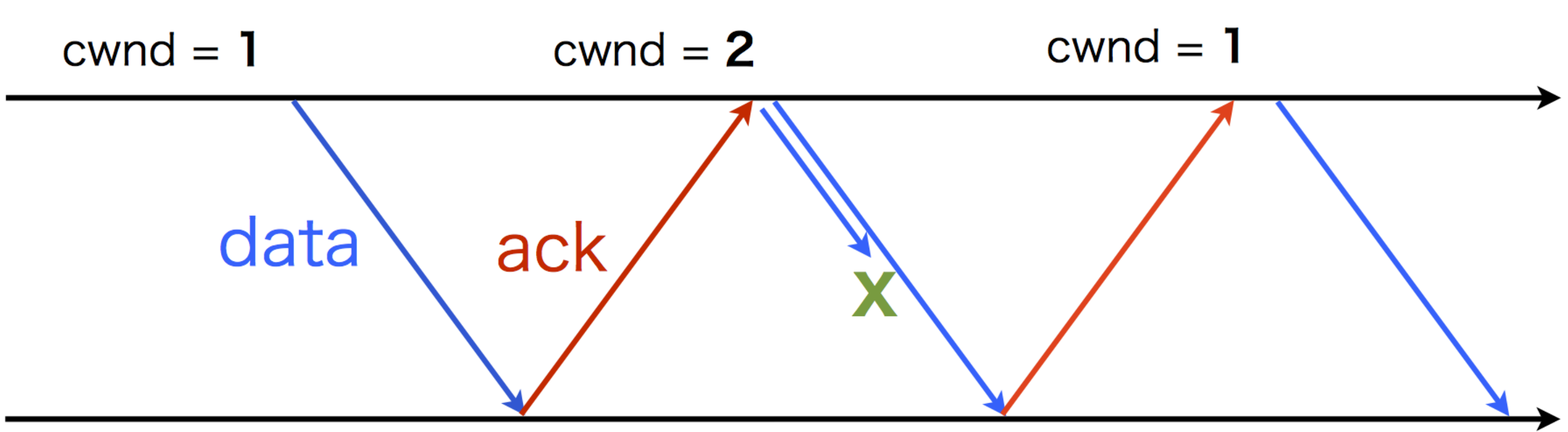}
\caption{{\bf Illustration of TCP packets and ACKs.} The cwnd increases gradually as an ACK is received and decreases otherwise.
\label{fig:cwnd_mechanism}}
\end{figure}

Sending packets one at a time with a constant speed and periodicity is inefficient. It is better to send as many packets as possible simultaneously without causing significant congestion. TCP's cwnd schema controls the number of packets that can be sent at one time. To achieve improved performance, the cwnd needs to be set to a large size; however, when the cwnd is too large, it creates congestion with other packets, resulting in multiple packet drops and a decrease in the overall throughput. A simple idea is to increase the cwnd size when it receives an ACK and to decrease it when a packet drop is detected (see Figure~\ref{fig:cwnd_mechanism}).

\subsection{TCP algorithms}
A number of different algorithms can be used to update the cwnd size. Currently, a realistic transport protocol based on the additive increase/multiplicative decrease algorithm is the most widely used across the Internet. In particular, the Tahoe, Reno, and NewReno protocols have been studied extensively using ns-2, a commonly used Internet simulator\footnote{The ns-2 network simulator is available at http://www.isi.edu/nsnam/ns/.}. Of these protocols, we used the NewReno algorithm for the experiments because it is the most recently developed and because it is widely used\footnote{TCP NewReno http://tools.ietf.org/html/rfc3782.}. Due to congestion, the NewReno algorithm increases the cwnd exponentially until the first packet drop occurs. In the event of drop, the cwnd is set to half of its value and then starts to increase again. In practice, when the sender does not receive an ACK for a certain period, a node resets the cwnd size to one and starts to increase again.

\begin{figure}[!ht]
\begin{center}
\includegraphics[width=0.80\textwidth]{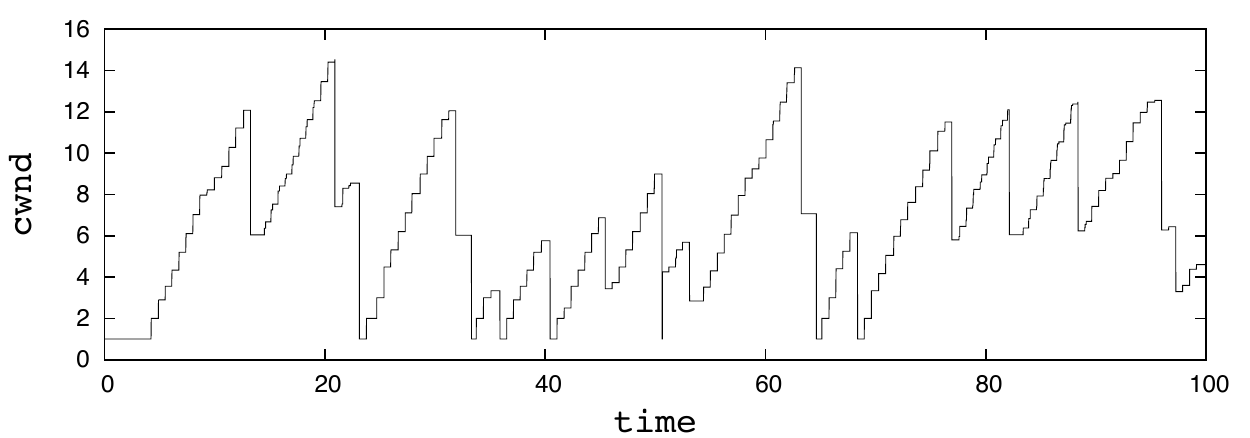}
\end{center}
\caption{ {\bf Example of cwnd dynamics on ns-2.}  The cwnd size increases quickly as it starts to send packets until it reaches a predefined threshold or a packet drop occurs, in which case the cwnd size decreases.
\label{fig:cwnd_example}}
\end{figure}

An example of cwnd dynamics using ns-2 is shown in Figure~\ref{fig:cwnd_example}. The cwnd's size increases quickly as it starts to send packets until it reaches a predefined threshold or a packet drop occurs, in which case, the cwnd size decreases. Next, the node starts sending packets again, and the cwnd increases in size incrementally as it continues to receive ACKs until another drop occurs. This simple feedback mechanism ensures that packets are sent to their respective destinations. The robustness and adaptability of PSNs is attributed to the inherent dynamics of cwnd.

\subsection{Simulations on ns-2}
We used ns-2 simulations to determine how various temporal cwnd
developments are organized. We constructed a simple network topology
comprising 30 nodes. The nodes are lined up on one dimensional line
and each node is connected to its right and left adjacent nodes and
acting as a relay node. We labeled these nodes in order from {\tt 0}
to {\tt 29}. In this network, we set 30 different flows specifying
source and destination nodes. The $k$th flow sends packets from node
$k$, and the destination node is set as node $(N - k - 1)$. For
example, the flow {\tt 0} sends packets from the source node {\tt 0}
to the destination node {\tt 29}, and the flow {\tt 29} sends packets
from {\tt 29} to the destination node {\tt 0}. The ns-2 simulator
periodically feeds input packets to each node. The input packet ratios
(which we call \textit{duty ratio}) vary from 0 to 1. When a duty
ratio $x$ is set as $x \leq 1$, each source node sends packets for $x$
seconds and then rests for the next $(1 - x)\times$ seconds in every
cycle. The simulator was run for 10,000 seconds.

In most works on PSNs, only a few nodes are used to simulate the PSN~\citep{tcp_chaotic, tcp_nonlinear, anna_gilbert, zhang_tcp, misako_takayasu}. We employed a topology comprising 30 nodes in order to create various types of bottlenecks within the network, since overall network performance is determined by bottlenecks due to packet congestion. In this setting, the network has nodes with different activity levels for packet processing. The busiest nodes are the centered nodes ({\tt 14} and {\tt 15}), which deal with multiple packets from both directions because all flows have to pass through them. The least busy nodes are on the boundaries ({\tt 0} and {\tt 29}). The longest flows are flow {\tt 0} and flow {\tt 29}, both of which cross the entire network, whereas the shortest are flow {\tt 14} and flow {\tt 15}. The TCP algorithms attempt to resolve congestion by regulating cwnd sizes.

\subsection{Dynamics of packet flow and cwnd}
\begin{figure}[!ht]
\begin{center}
\subfigure[duty ratio = 0.10]{
\includegraphics[width=0.80\textwidth, bb=0 0 3000 1050]{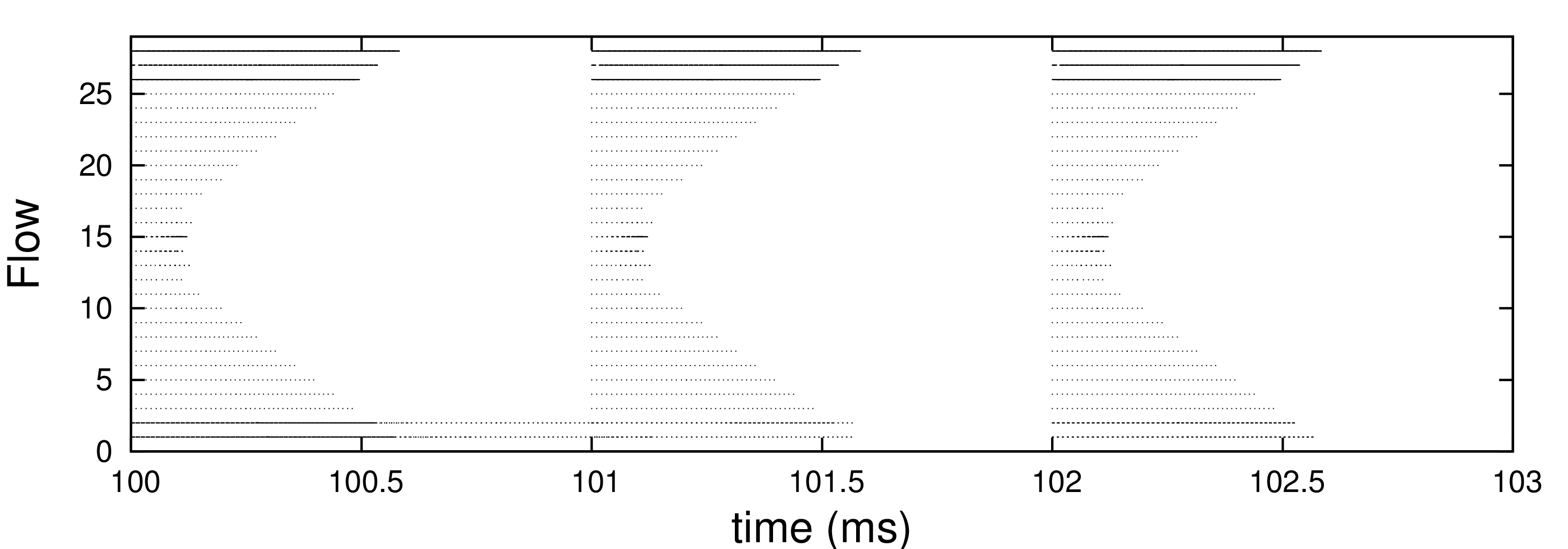}}
\subfigure[duty ratio = 0.50]{
\includegraphics[width=0.80\textwidth, , bb=0 0 3000 1050]{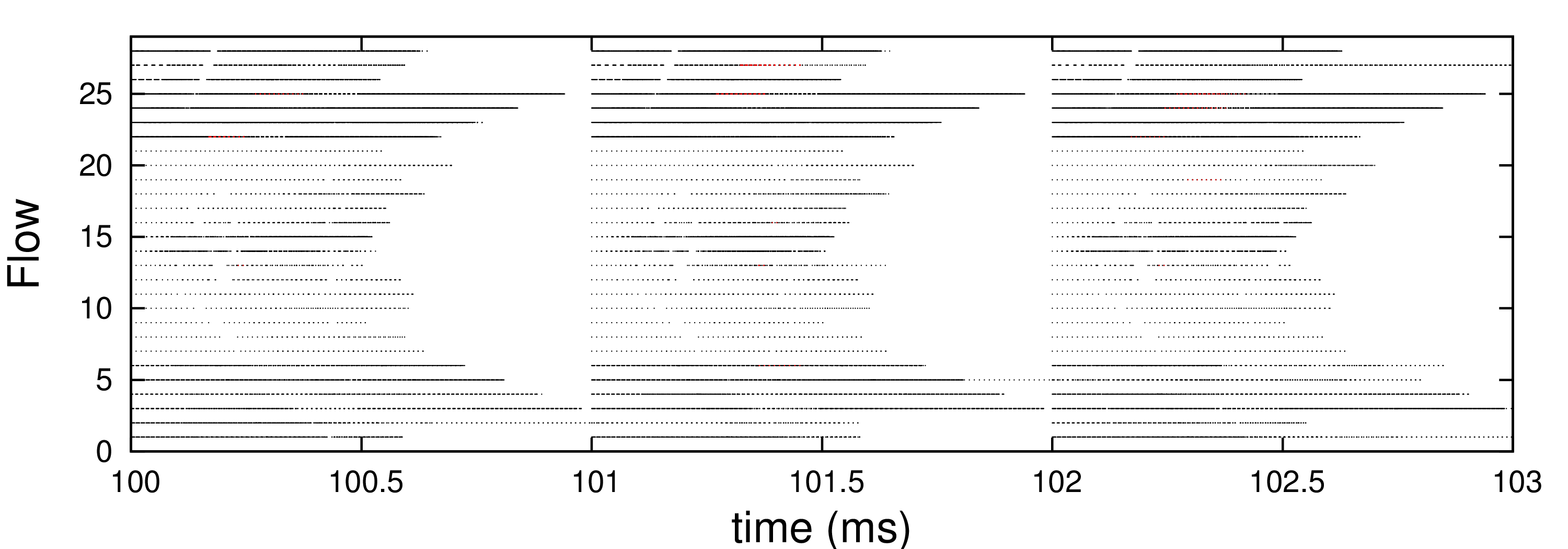}}
\subfigure[duty ratio = 0.80]{
\includegraphics[width=0.80\textwidth, bb=0 0 3000 1050]{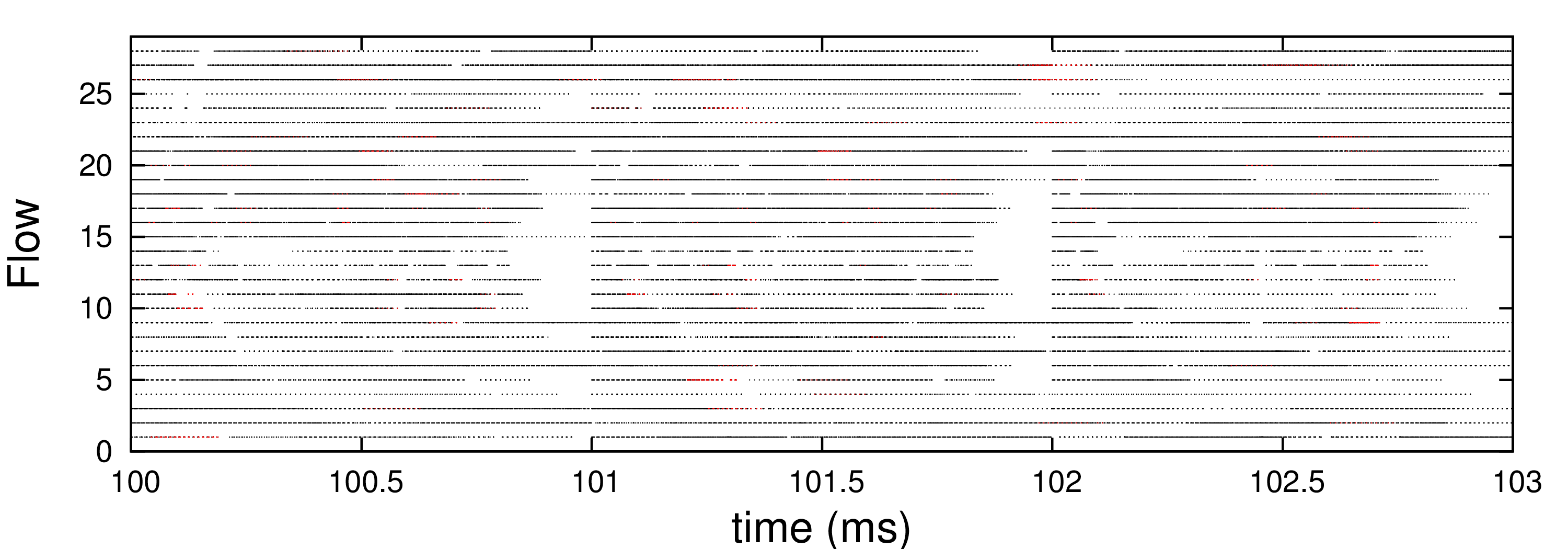}}
\end{center}
\caption{ {\bf Temporal development in packet transmissions for all 30 flows in the network.}  The source node of each flow is plotted on the y-axis, and when the packets of each flow are transmitted from one of the relay nodes, the packet is marked at the corresponding time step (x-axis). Thus, each dot corresponds to the transmission event of a packet in the flow, and dropped packets are shown in red. When the duty is low (top), the plot is a V-shaped curve, with the shortest paths at the middle nodes representing the minimum packet transmission time and the longest paths at the peripheral nodes representing the maximum packet transmission time. When the duty ratio is higher, the shape has two more peaks (middle). This corresponds to the origin of congestion. For higher duty rations, a moth-eaten shape with many drop events is observed (bottom).
\label{fig:flow_send}}
\end{figure}

Given the above setup, we first observed the behaviors of each flow with respect to the duty ratio. If the topology is set homogeneously and the packets do not overflow the buffer capacity, the packet transmission time for each flow should be proportional to the corresponding flow length. However, because the network has a bottleneck, packet flow behavior is more complicated. Figure~\ref{fig:flow_send} shows the temporal development of packet sending when the input duty ratio is varied as $x = 0.10, 0.50$, and $0.80$ for all 30 flows. When $x = 0.10$, the packet transmission times of the 30 flows form a "V" shape, shown in Figure~\ref{fig:flow_send}. This is because flows {\tt 14} and {\tt 15}, which pass through the centered nodes, traverse the shortest path, while flows {\tt 0} and {\tt 29} traverse the longest path. As the duty ratio increases, the packet sending times do not retain this "V" shape; rather, they begin to show a more complex shape due to congestion. Packet drops are colored in red in the figure, which shows that the frequency of drops increases with an increase in the input duty ratio, thus indicating the various types of temporal packet transmission dynamics at play.

\begin{figure}[!ht]
\centering
\subfigure[Flow = 0]{
\includegraphics[width=0.48\textwidth, bb=0 0 3000 2100]{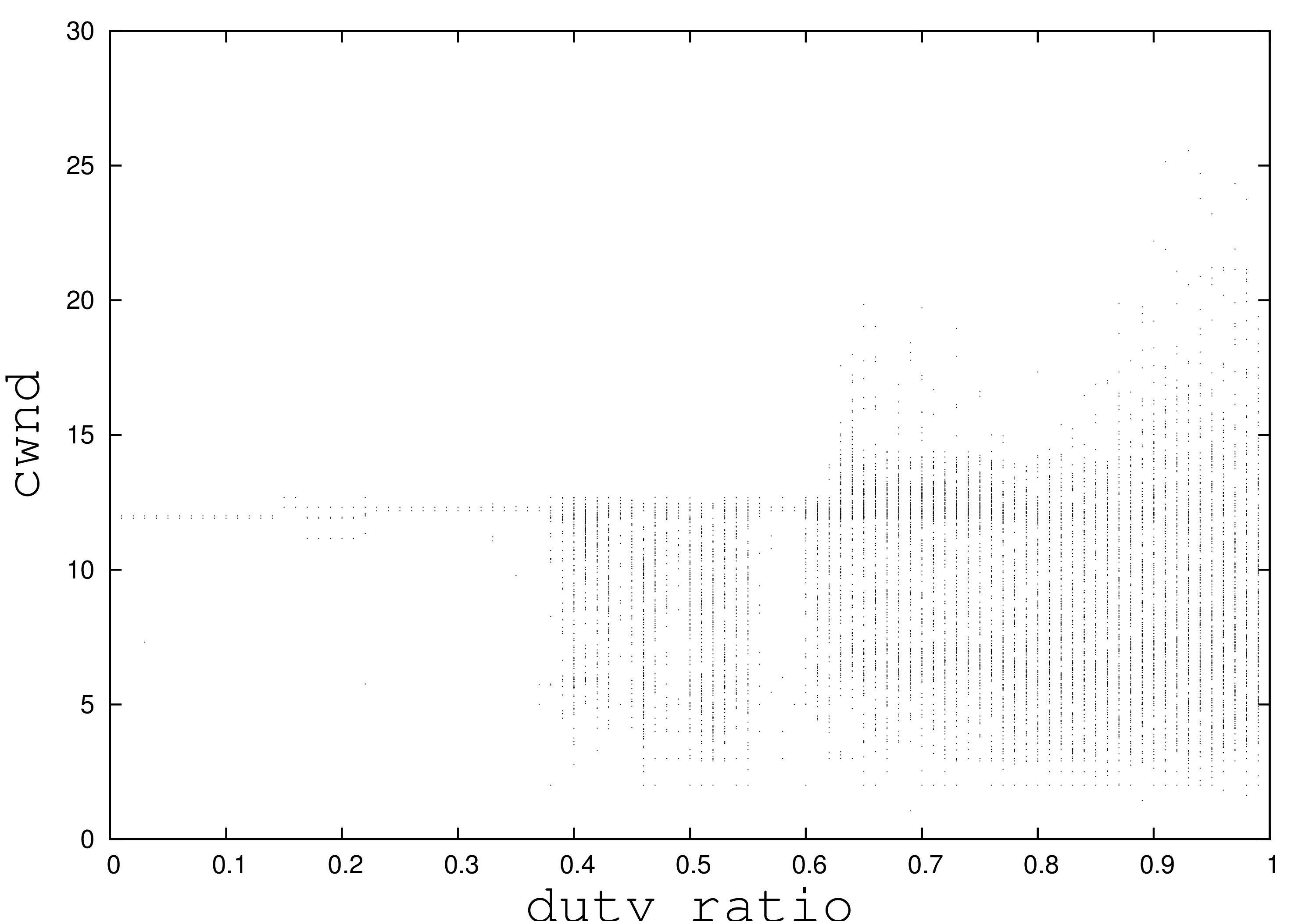}}
\subfigure[Flow = 5]{
\includegraphics[width=0.48\textwidth, bb=0 0 3000 2100]{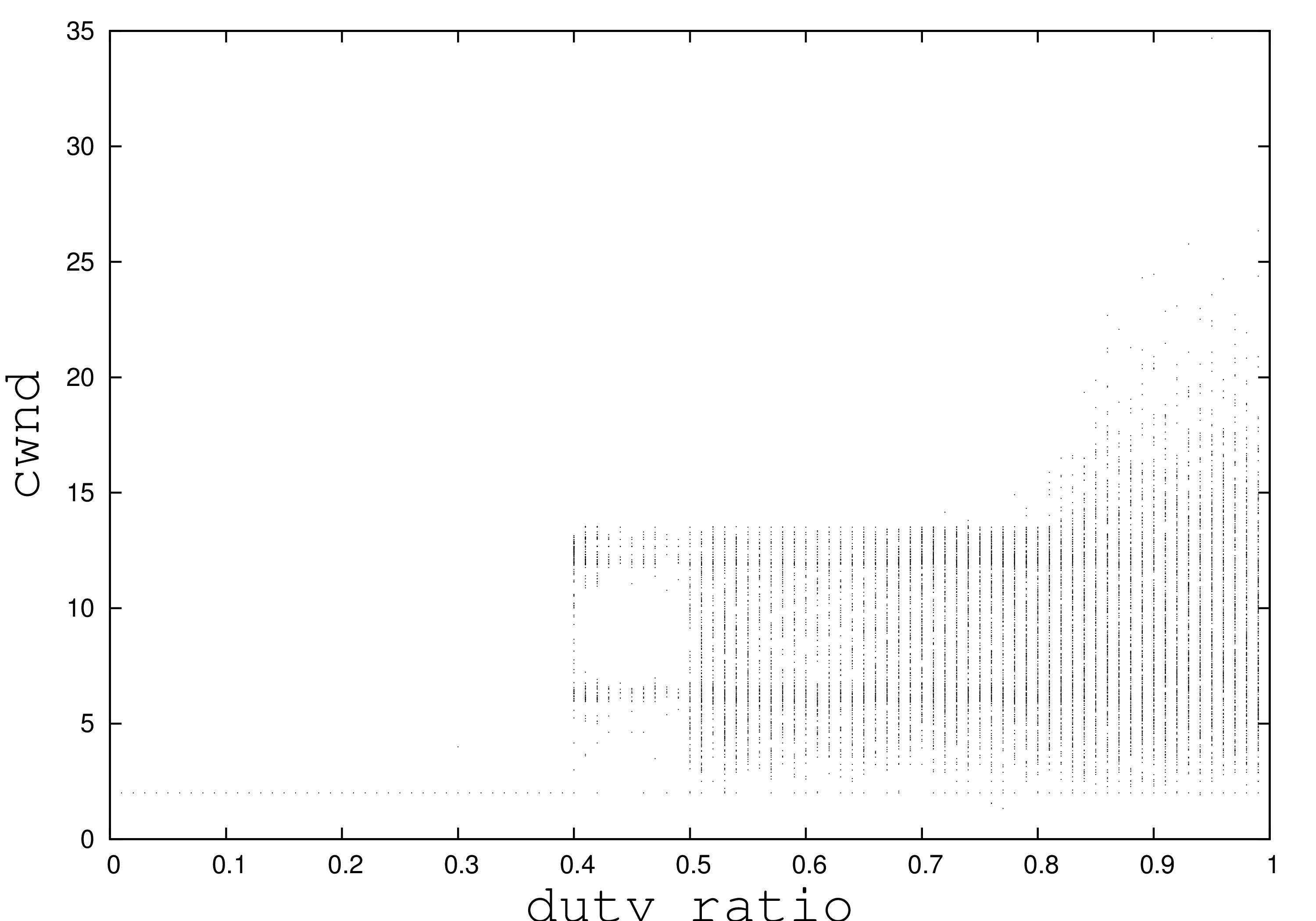}}
\subfigure[Flow = 15]{
\includegraphics[width=0.48\textwidth, bb=0 0 3000 2100]{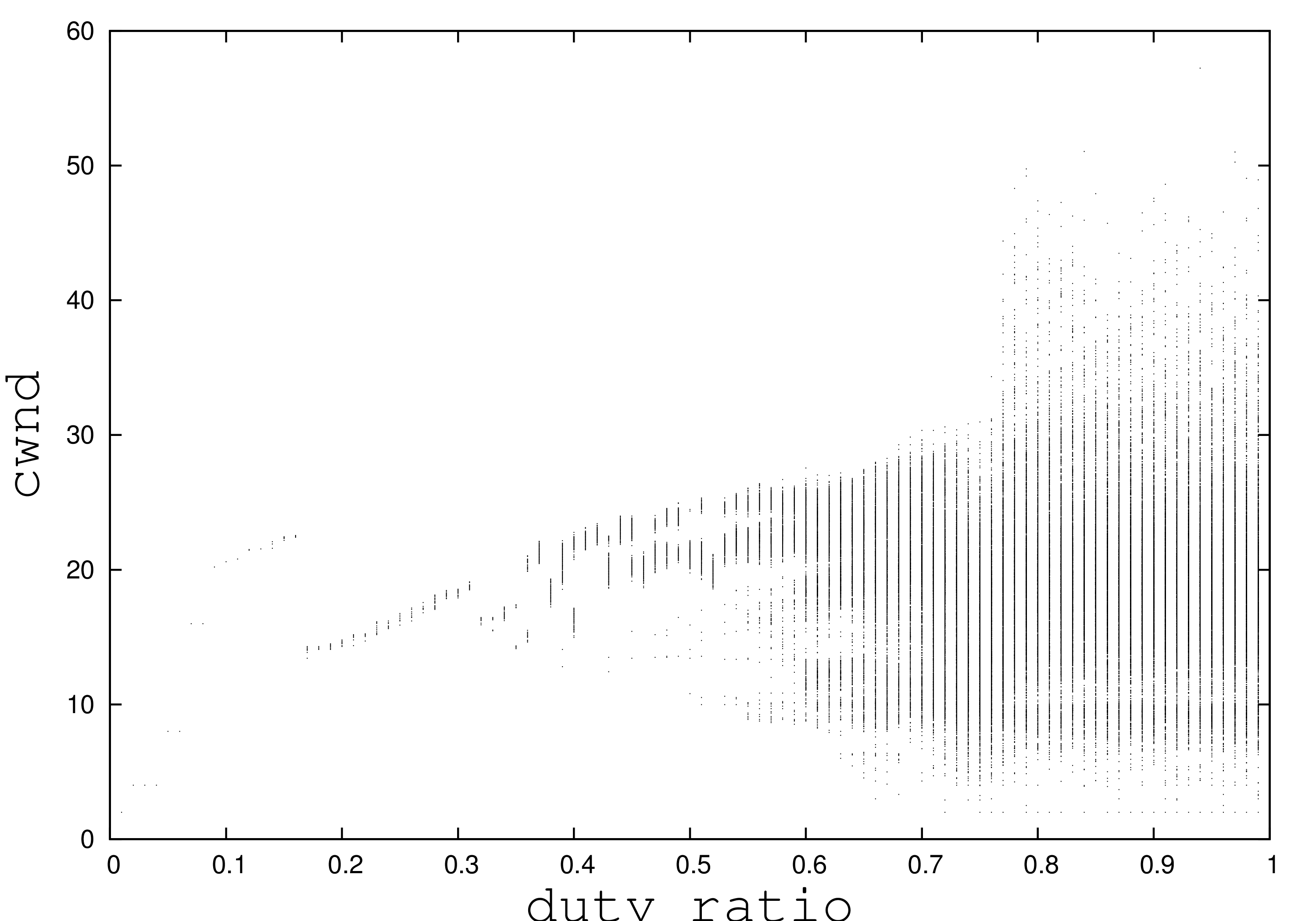}}
\subfigure[Flow = 20]{
\includegraphics[width=0.48\textwidth, bb=0 0 3000 2100]{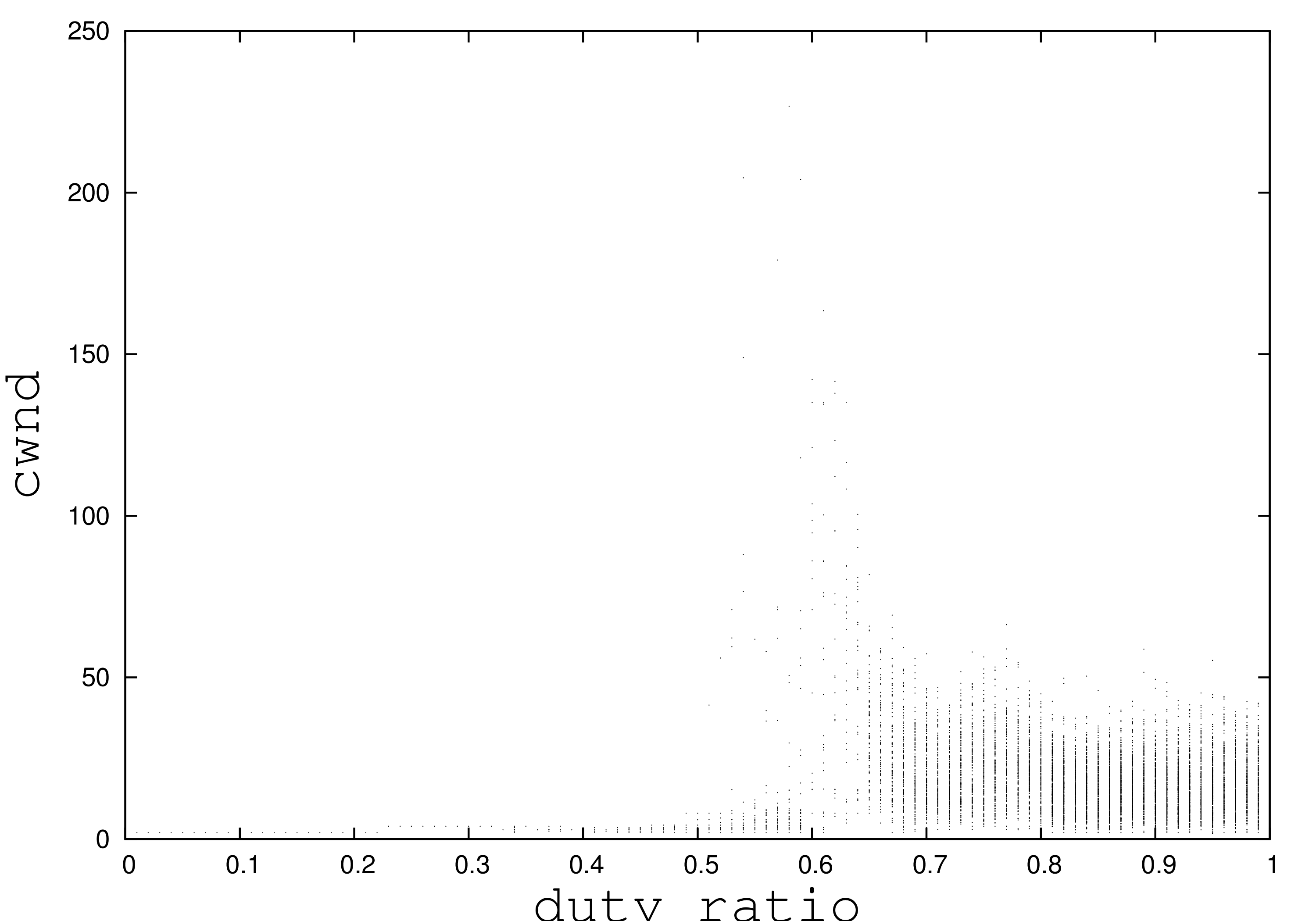}}
\caption{ {\bf cwnd bifurcation diagrams drawn by taking local peaks of cwnd time series for flows {\tt 0}, {\tt 5}, {\tt 15}, and {\tt 20} with respect to duty ratio.} Changes in the cwnd dynamics from periodic to chaotic are shown.
\label{fig:cwnd_bifurcation}}
\end{figure}

The complex dynamics of packet transmission are controlled by the cwnd; thus, we look at the underlying temporal dynamics of the cwnd. Figure~\ref{fig:cwnd_bifurcation} shows as examples the bifurcation diagrams of flows {\tt 0}, {\tt 5}, {\tt 15}, and {\tt 20} in relation to the duty ratio. A bifurcation diagram is plotted by taking the local peaks of the cwnd time series. Notably, the cwnd temporal behaviors change from fixed points to periodic and, then, to chaotic as the number of packets to be transmitted increases~\footnote{We have estimated the Lyapunov exponent (the degree of instability) from the time series of cwnd showing the positive values.}.
This chaotic behavior is the source of the various types of temporal dynamics of packet transmission in each flow.  Chaotic dynamics and the bifurcation route to the chaotic state has been reported previously; however, it has only been reported in a low-dimensional system~\citep{tcp_chaotic,tcp_nonlinear}. Here, instead, we report the bifurcation patterns of a spatially extended system.

\section{PSN as Dynamical Systems}
The simple observation of the previous section implies that a PSN is characterized by complex, underlying cwnd dynamics that keep the system working robustly in the face of dynamic changes. In fact, the average number of successfully transmitted packets keeps increasing, despite congestion events. This is because the TCP algorithm is designed to maintain a stable average throughput~\citep{Low_1999}. We describe this mechanism in terms of the dynamics of PSN states and attractors. We re-fine the dynamic state of PSN by using the two largest components of PCA computed from the cwnd time series (will be defined in the following section). 
We investigate how the states evolve and create attractors over time by drawing a transition diagram of the states and by discussing their relationship with the throughput. In this paper, we define \textit{throughput} as the overall percentage of successfully sent packets, or 
$$
\mbox{throughput} = \frac{(\mbox{number of sent packets} - \mbox{number of drops})}{\mbox{number of sent packets}}
$$

\subsection{Perturbation}
\begin{figure}[!ht]
\centering
\subfigure[Flow = 2, duty = 0.20]{
\includegraphics[width=0.60\textwidth]{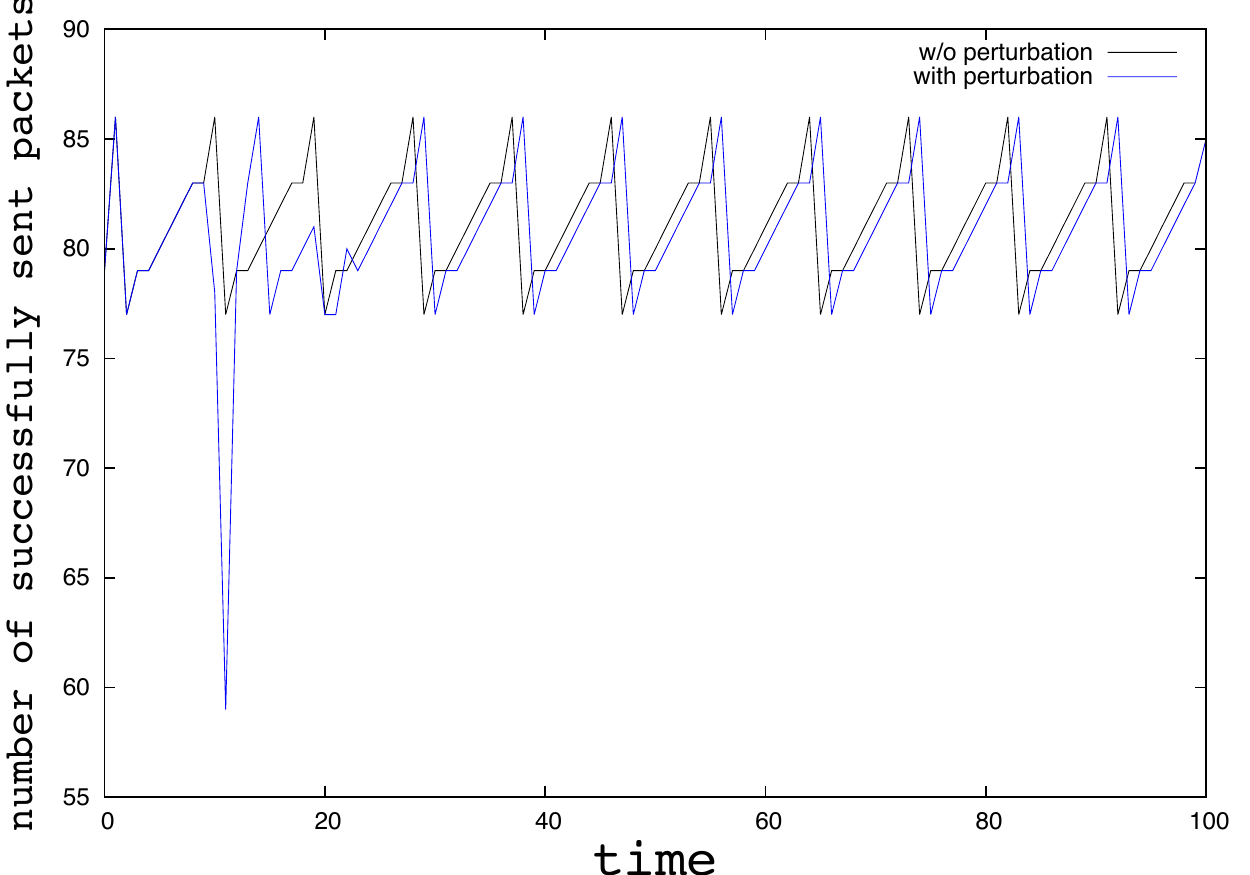}}
\subfigure[Flow = 2, duty = 0.40]{
\includegraphics[width=0.60\textwidth]{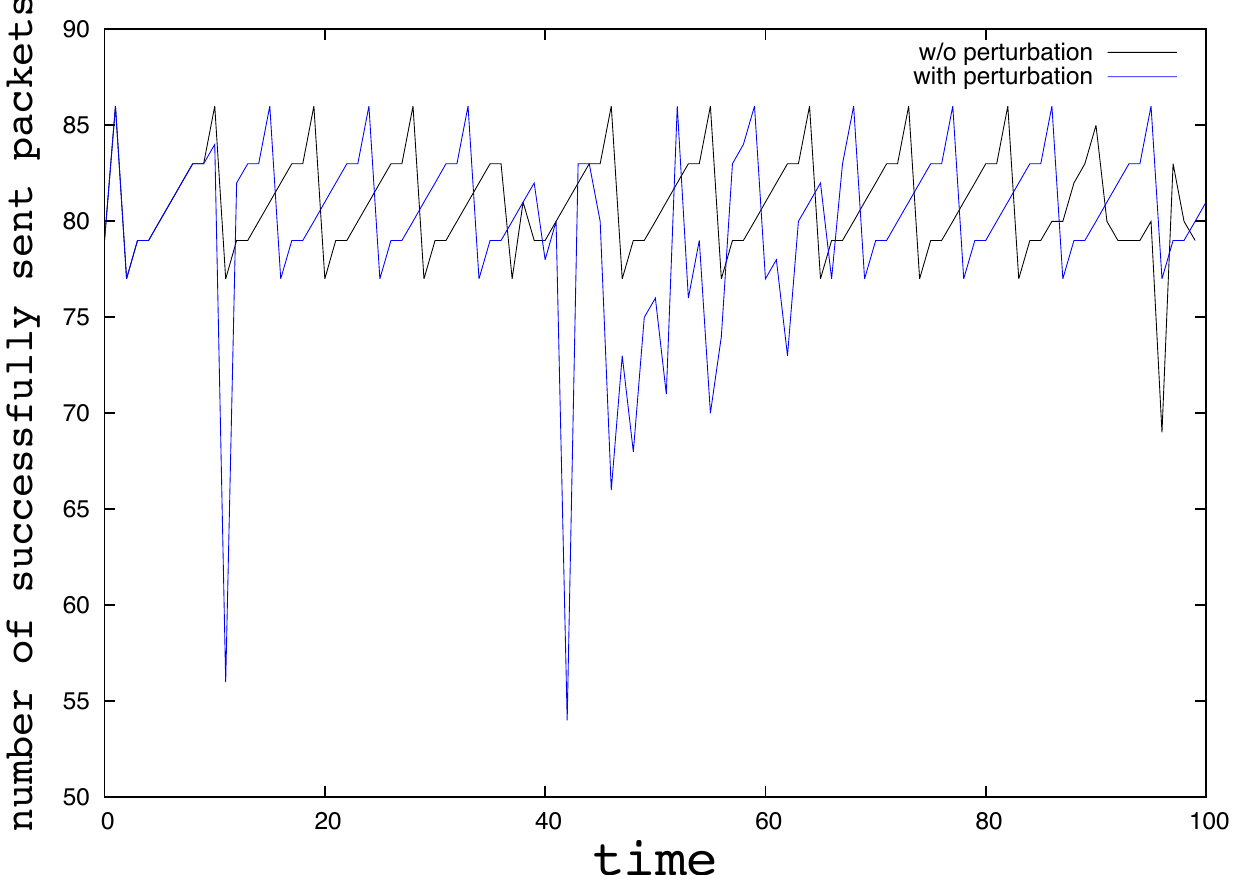}}
\subfigure[Flow = 2, duty = 0.60]{
\includegraphics[width=0.60\textwidth]{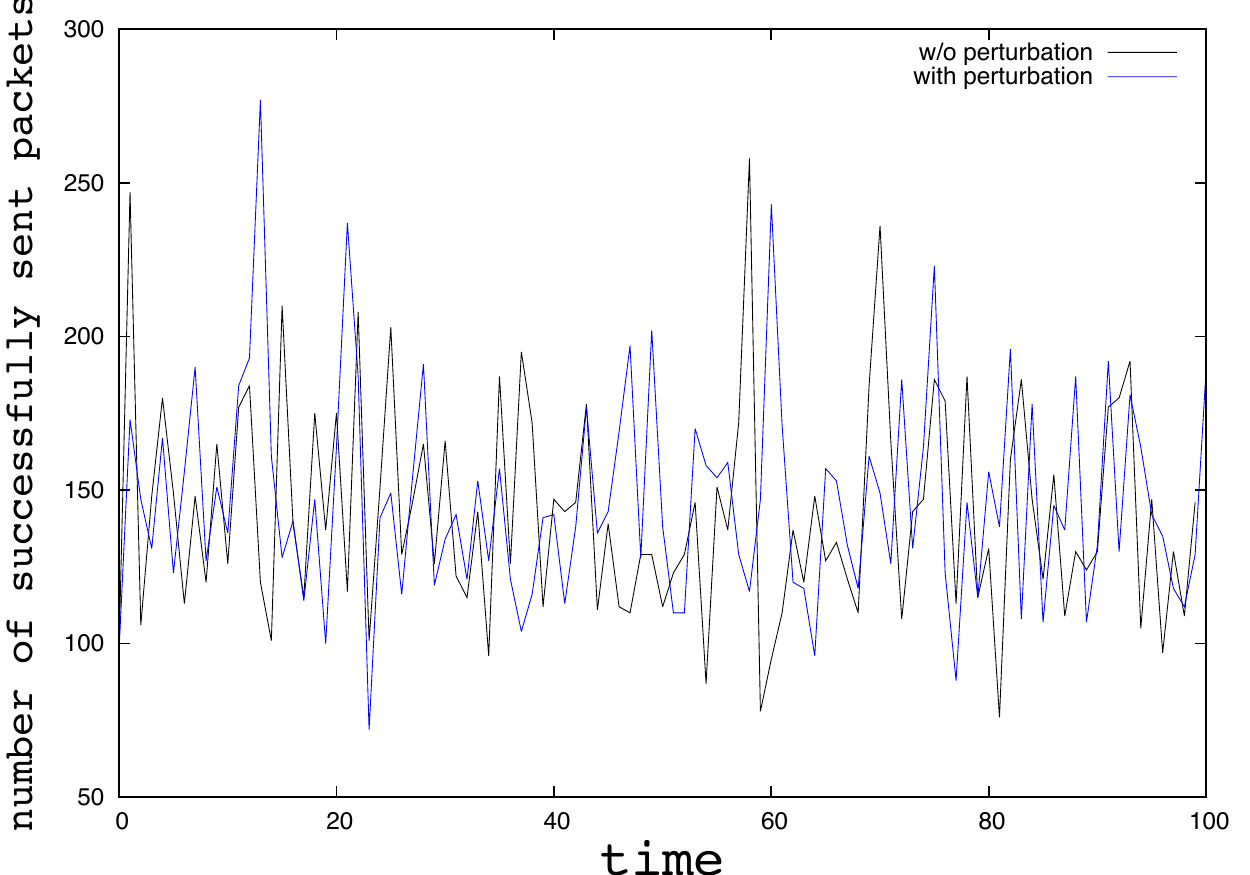}}
\caption{ {\bf Temporal development of throughputs with or without perturbation for flow {\tt 2} with duty ratios of $x = 0.20, 0.40$, and $0.60$.} Additional packets were added for a duration of 10 seconds each at intervals of 100 seconds in order to perturb the system. Throughput without perturbation is indicated by a black line, and that throughput with perturbation is indicated by a blue line.
\label{fig:with_or_without}}
\end{figure}

Additional packets from node {\tt 0} to {\tt 29} were added for durations of 10 seconds each at intervals of 100 seconds in order to perturb the system. The total duration of each simulation was 10,000 seconds; thus, 99 perturbations were added to the system. The temporal development of the number of transmitted packets with and without perturbation is shown in Figure~\ref{fig:with_or_without}. When the duty ratio is small with little congestion ($x = 0.20$), the system dynamics restore the original behavior of sending packets, even after the addition of perturbation. As the duty increases ($x = 0.40$), the system dynamics take longer to return to the original behavior. Finally, with a further increase in the duty ($x = 0.60$), the dynamics become more complex, and it is difficult to determine whether the system remains within the same behavior. 

\subsection{A PSN state by principal component analysis}
To see the underlying dynamics behind the behavior changes due to the perturbation, we define a dynamical state of a PSN by projecting the cwnd time series onto a reduced-feature vector space using principal component analysis.

\begin{figure*}[!ht]
\begin{center}
\subfigure[duty = 0.20] {
  \includegraphics[width=0.24\textwidth,bb=0 0 300 300]{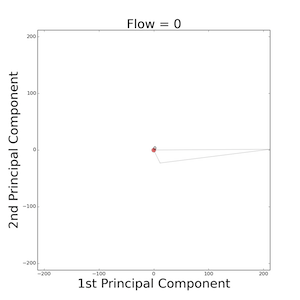}
  \includegraphics[width=0.24\textwidth,bb=0 0 300 300]{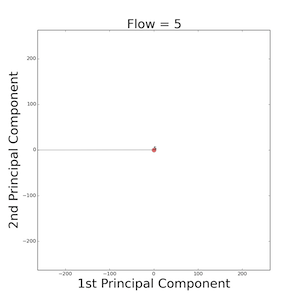}
  \includegraphics[width=0.24\textwidth,bb=0 0 300 300]{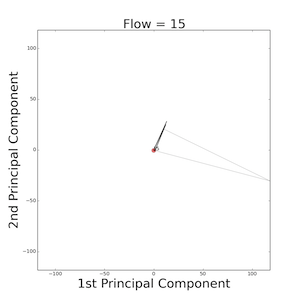}
  \includegraphics[width=0.24\textwidth,bb=0 0 300 300]{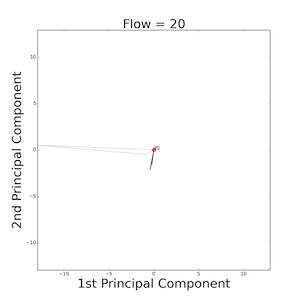}
}
\subfigure[duty = 0.40] {
  \includegraphics[width=0.24\textwidth,bb=0 0 300 300]{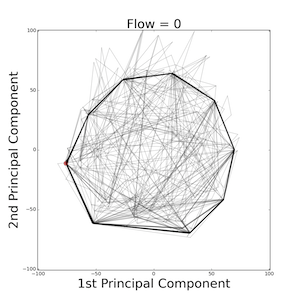}
  \includegraphics[width=0.24\textwidth,bb=0 0 300 300]{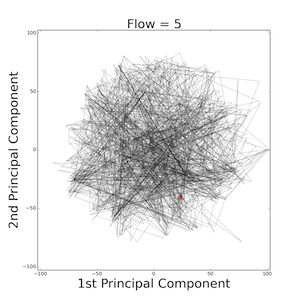}
  \includegraphics[width=0.24\textwidth,bb=0 0 300 300]{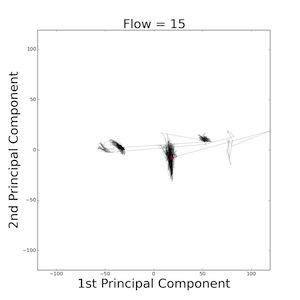}
  \includegraphics[width=0.24\textwidth,bb=0 0 300 300]{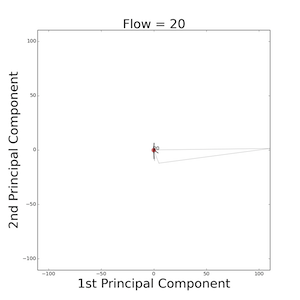}
}
\subfigure[duty = 0.60] {
  \includegraphics[width=0.24\textwidth,bb=0 0 300 300]{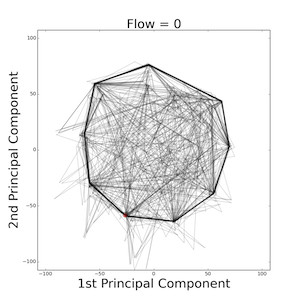}
  \includegraphics[width=0.24\textwidth,bb=0 0 300 300]{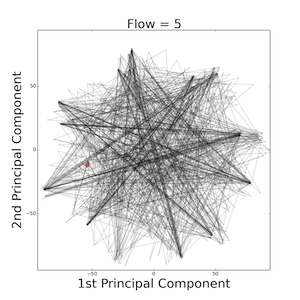}
  \includegraphics[width=0.24\textwidth,bb=0 0 300 300]{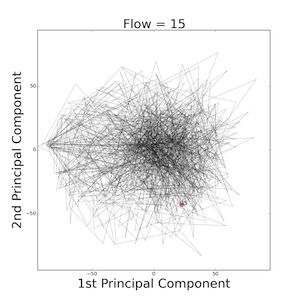}
  \includegraphics[width=0.24\textwidth,bb=0 0 300 300]{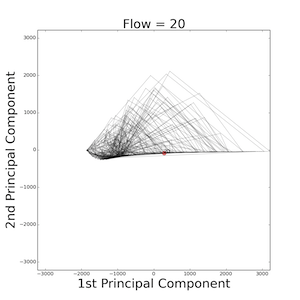}
}
\caption{ {\bf Examples of trajectories for the principal component space for flows {\tt 0}, {\tt 5}, {\tt 15}, and {\tt 20}.} The first-largest principal component on the x-axis and the second-largest principal component on the y-axis. The upper panels show traces of situations in which the input is relatively low ($x = 0.20$), the middle panels correspond to higher inputs ($x = 0.40$) and the lower panels correspond to even higher inputs   ($x = 0.60$).
 \label{fig:pca_space}}
\end{center}
\end{figure*}

We took the following steps to obtain dynamical states: 1) Divide the time series of each node into windows of $T$ seconds ($T = 10$ in this experiment, thus creating 1,000 windows over a simulation duration of 10,000 seconds). Each window has 1,000 steps because the cwnd size is recorded every 0.01 seconds. This creates a matrix of 1,000 steps $\times$ 1,000 windows. 2) Apply principal component analysis to the matrix and obtain two-dimensional feature vectors for each window using the largest and the second-largest eigenvalues. We plotted each projected window onto this principal component space and drew trajectories of the coordinates using the first-largest principal component on the x-axis and the second-largest principal component on the y-axis. A few examples are shown in Figure~\ref{fig:pca_space}. The upper panels show traces of situations in which the input was relatively low ($x = 0.20$) (i.e., for flows {\tt 0}, {\tt 5}, {\tt 15}, and {\tt 20}), the middle panel correspond to higher inputs ($x = 40$), while the lower panels correspond to even higher inputs ($x = 0.60$). We can observe attractor self-organization in the state transitions. For example, in the flow {\tt 0}, with a duty ratio $x = 0.40$, we can clearly observe a strong limit cycle in period nine, with occasional transitions to other states. Flow {\tt 5} has a greater number of transitional states, and flow {\tt 15} has fewer transitional states, forming only a few attractors and transitioning among them. Flow {\tt 20} transitions among even fewer states. As the input increases - for example, to the duty ratio $x = 0.60$ - the presence of attractors becomes less obvious; the flow starts to create a greater number of states and transition among them.

\subsection{Transition diagram}
\begin{figure*}[!ht]
\begin{center}
\subfigure[Flow = 0] {
\includegraphics[height=0.30\textwidth,bb=0 0 347 445]{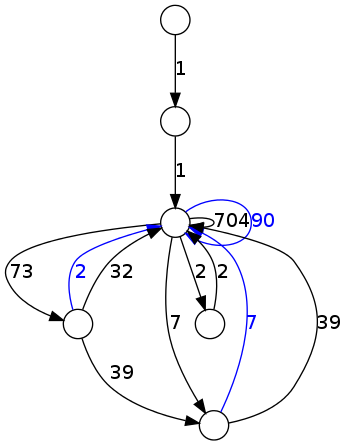}}
\subfigure[Flow = 5] {
\includegraphics[height=0.30\textwidth,bb=0 0 255 344]{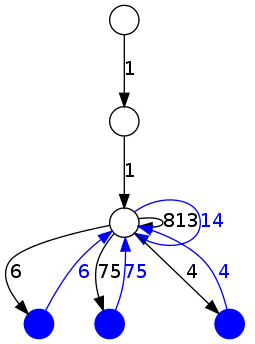}}
\subfigure[Flow = 15] {
\includegraphics[height=0.30\textwidth,bb=0 0 431 851]{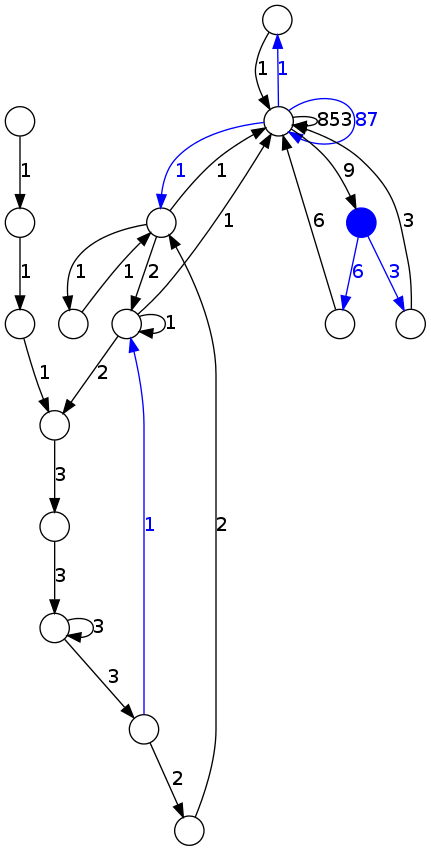}}
\subfigure[Flow = 20] {
\includegraphics[height=0.30\textwidth,bb=0 0 314 344]{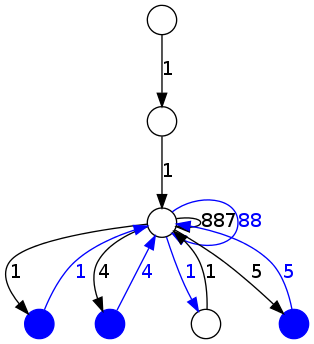}}

\subfigure[Flow = 0] {
\includegraphics[width=0.24\textwidth,bb=0 0 804 1256]{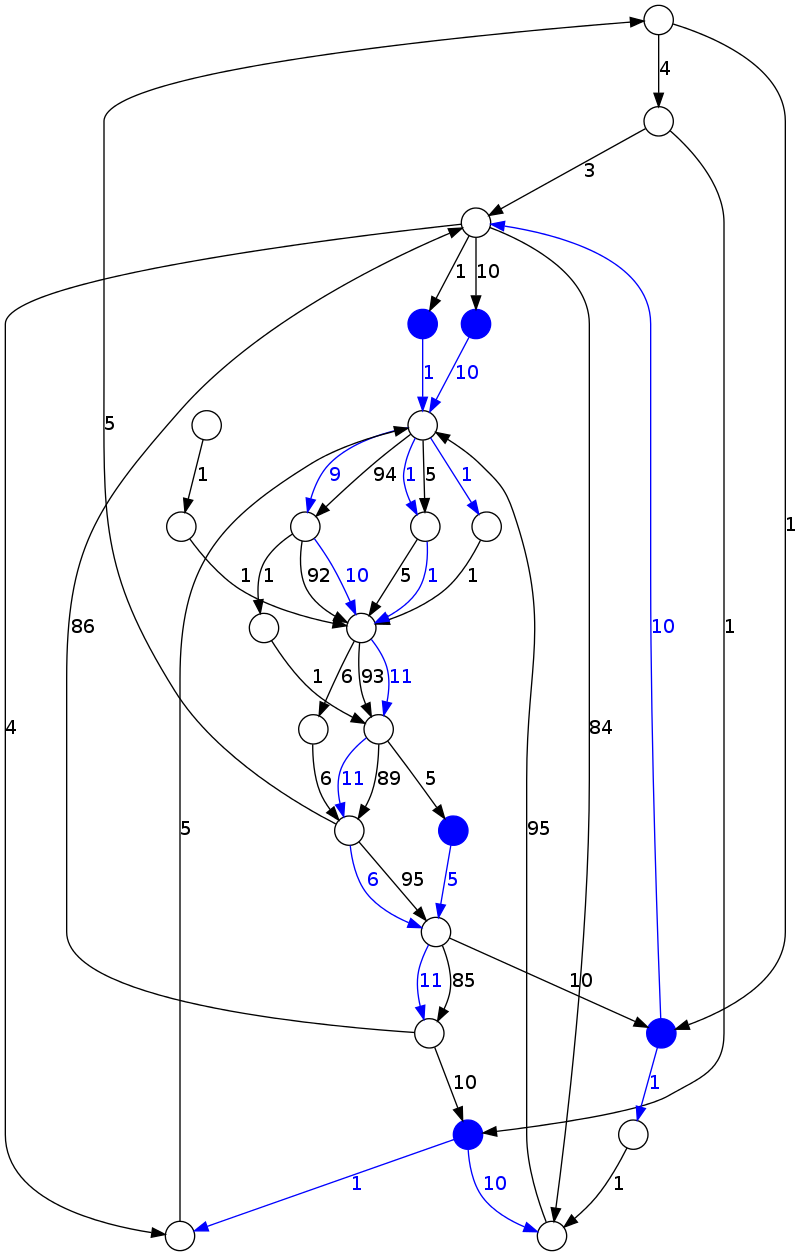}}
\subfigure[Flow = 5] {
\includegraphics[width=0.24\textwidth,bb=0 0 1989 3219]{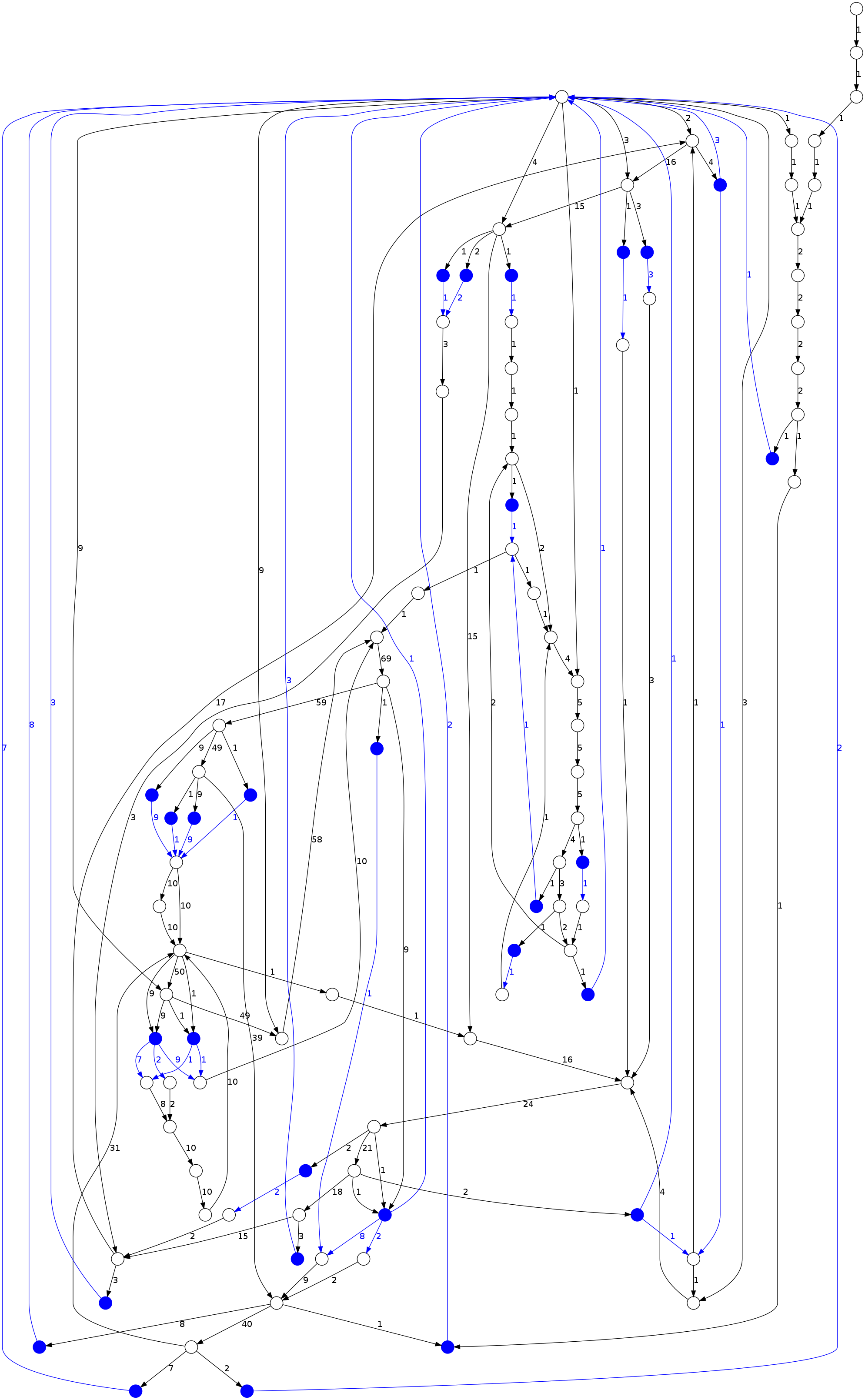}}
\subfigure[Flow = 15] {
\includegraphics[width=0.24\textwidth,bb=0 0 1316 3315]{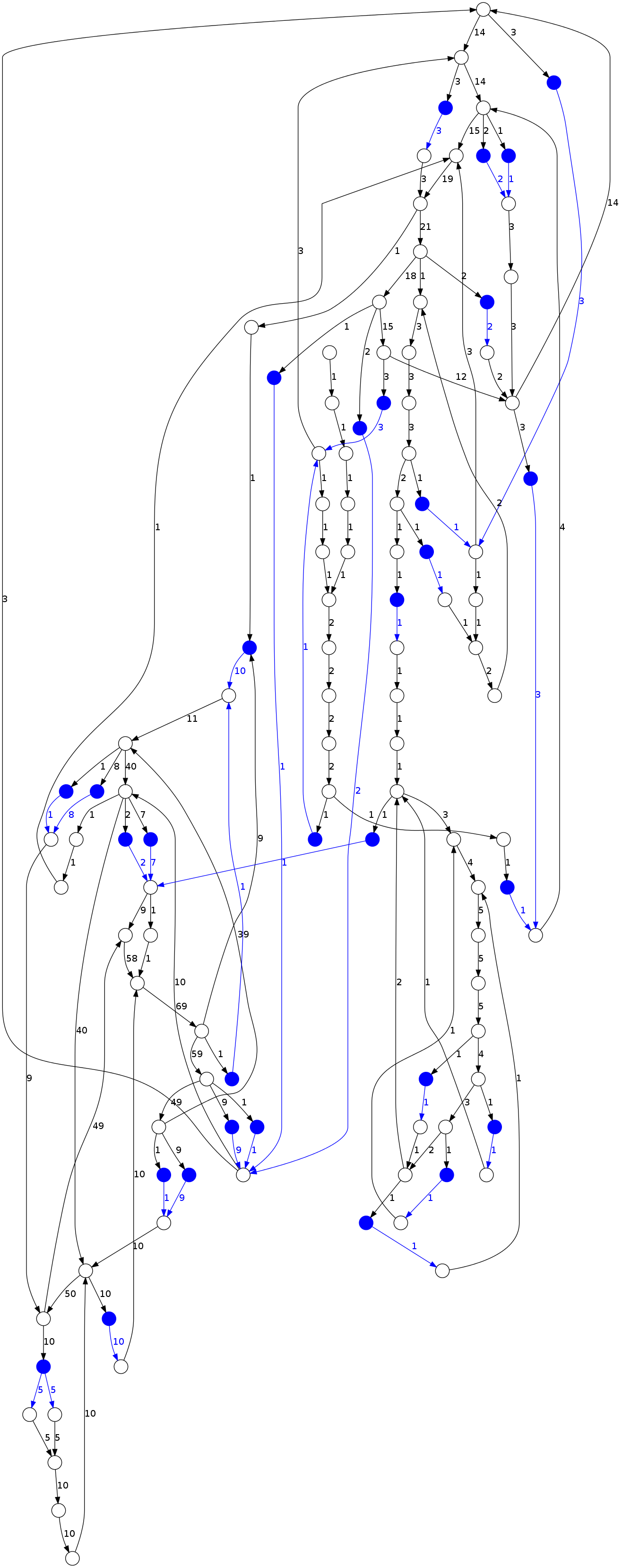}}
\subfigure[Flow = 20] {
\includegraphics[width=0.24\textwidth,bb=0 0 1531 243]{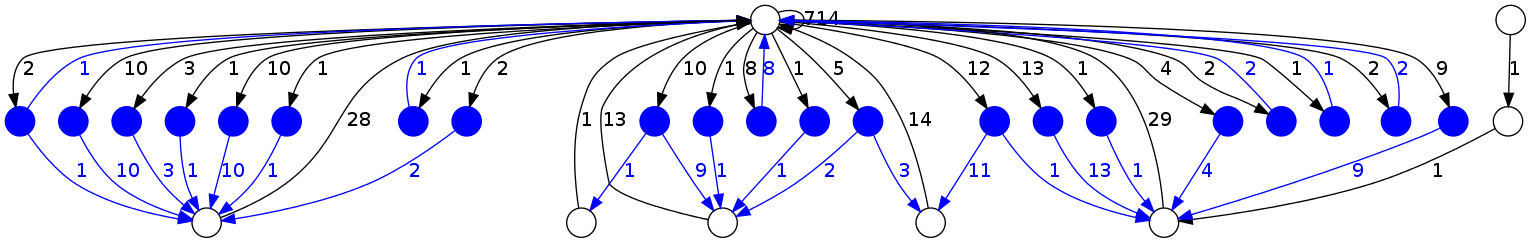}}
\end{center}
\caption{ {\bf Examples of state transition diagrams for flows {\tt 0}, {\tt 5}, {\tt 15}, and {\tt 20} with the duty ratio of $x = 0.20$ (top) and $x = 0.30$ (bottom).} The nodes colored in blue depict the state newly created by a perturbation, while the edges colored in blue depict transitions by perturbation. The value on the edge depicts the number of transition occurrences. 
 \label{fig:att_network}}
\end{figure*}

Let us take a closer look at the attracting state change with and without perturbation. The dynamic state specified by the two largest principal components seems to capture an attracting state, such as a stable-fixed or a temporally oscillating state. Taking the $(x,y)$ coordinates as states, we create a state transition diagram by connecting successive states to determine whether the system is in an attracting state. Taking a state as a node and a transition as an edge, we draw a transition diagram for each node with a duty. 
If it comes back to the original state, it forms a self-loop. 
If the transition is made to a node that already exists, then the transition becomes an edge to that node. If the transition is made to a new state, then a new node is created and forms an edge to the node. A few representative examples of state transition diagrams are shown in Figure~\ref{fig:att_network} for flows {\tt 0}, {\tt 5}, {\tt 15}, and {\tt 20} with the duty ratios $x =0.20$ and $x=0.30$. The states created due to perturbation are colored in blue; other states are shown in black. The weight on the edge represents the number of transition events. 
These flows are examples of cases in which a perturbation created a new state and transitioned there, or created a new transition but quickly transitioned to an existing state and remained within that state.

\subsection{Stability of states and throughput}
\begin{figure}[!ht]
\centering
\subfigure[Flow = 0]{
\includegraphics[width=0.48\textwidth]{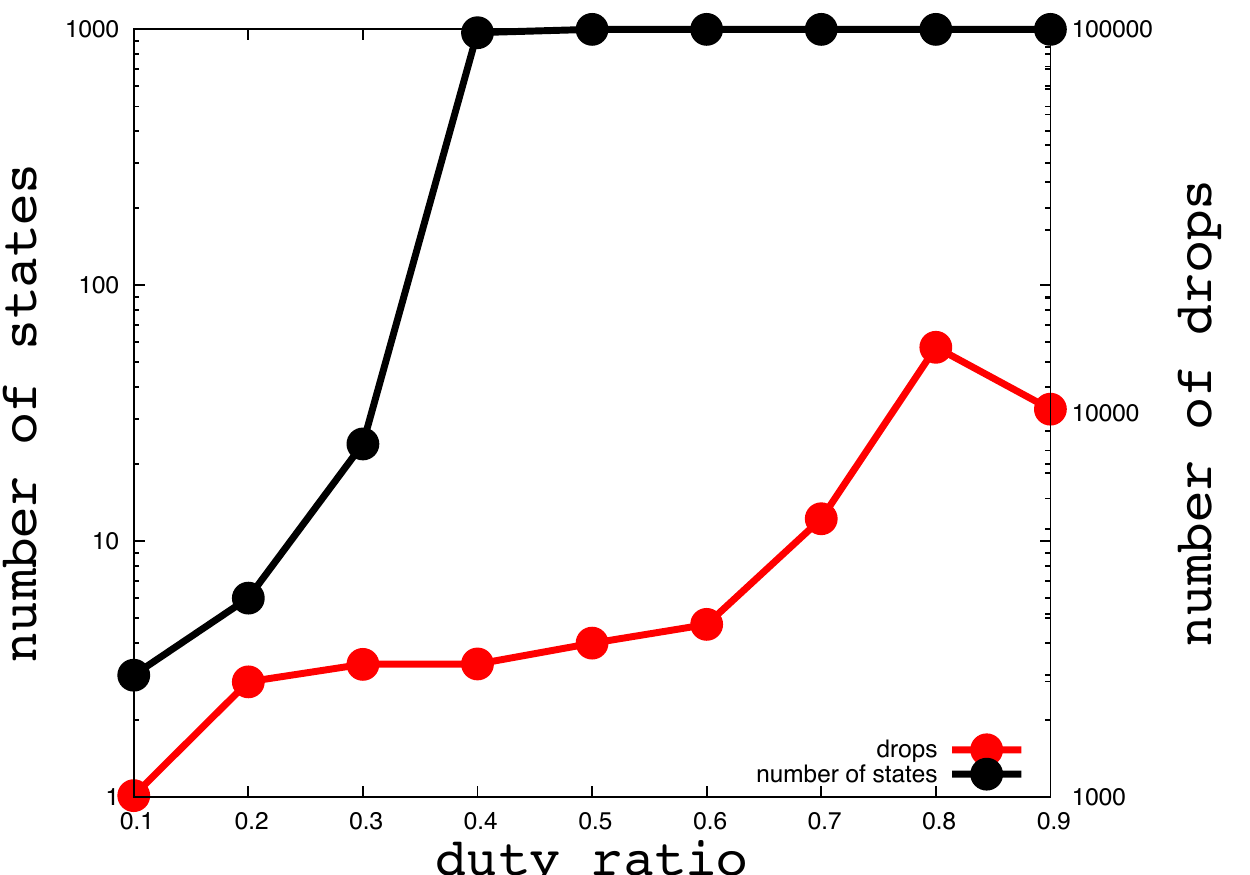}}
\subfigure[Flow = 5]{
\includegraphics[width=0.48\textwidth]{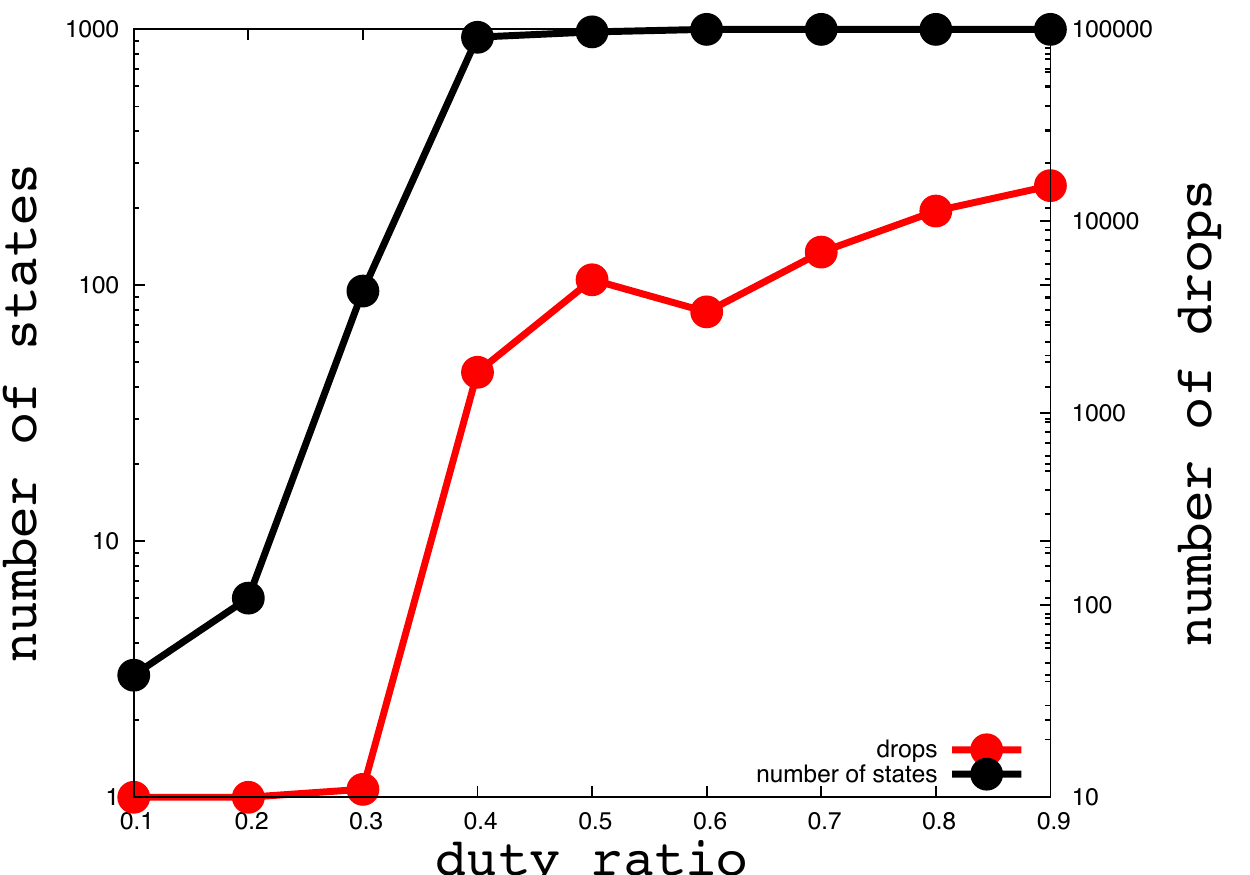}}
\subfigure[Flow = 15]{
\includegraphics[width=0.48\textwidth]{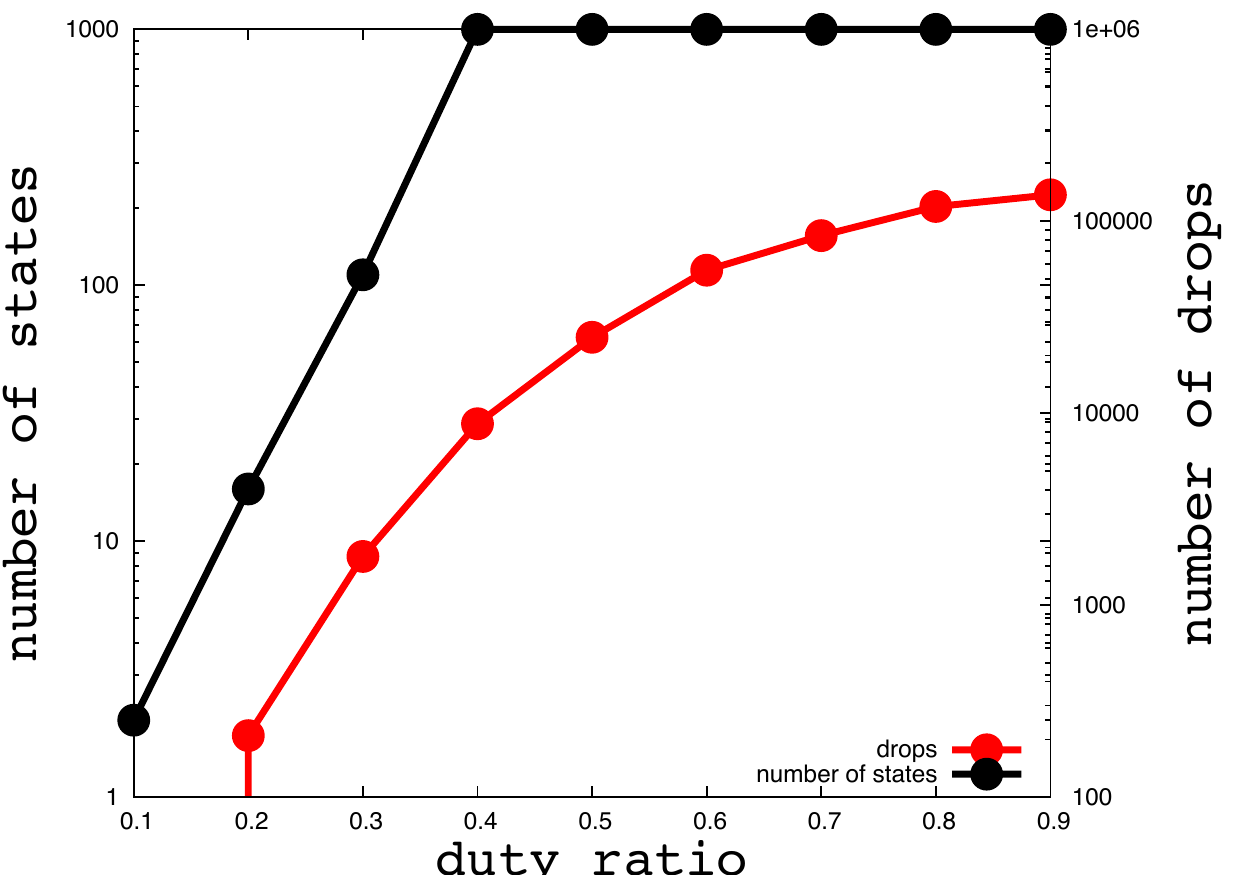}}
\subfigure[Flow = 20]{
\includegraphics[width=0.48\textwidth]{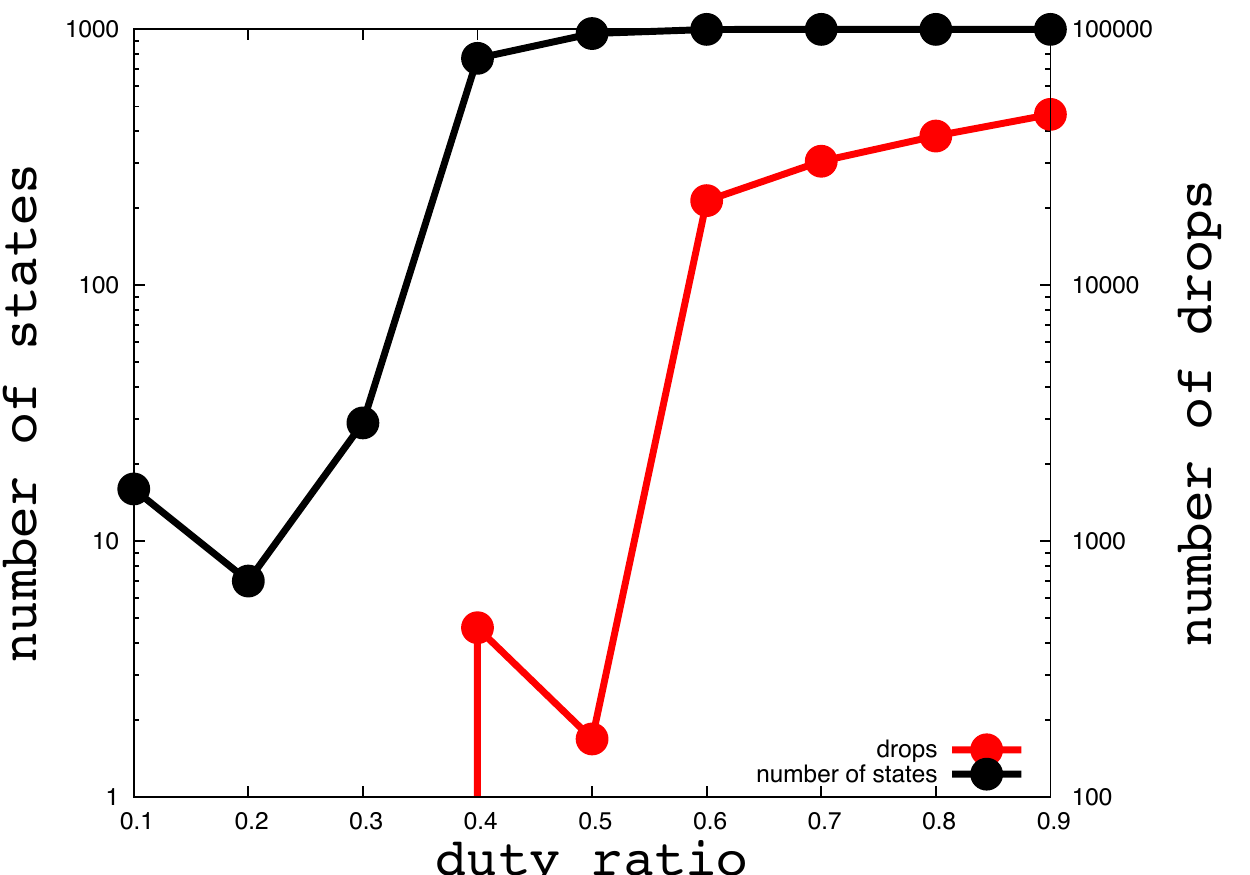}}
\caption{ {\bf Change in the number of states (in black) and the number of drops (in red) for flows {\tt 0}, {\tt 5}, {\tt 15}, and {\tt 20}.}}
\label{fig:num_states_drops}
\end{figure}

\begin{figure}[!ht]
\centering
\subfigure[Flow = 0]{
\includegraphics[width=0.48\textwidth]{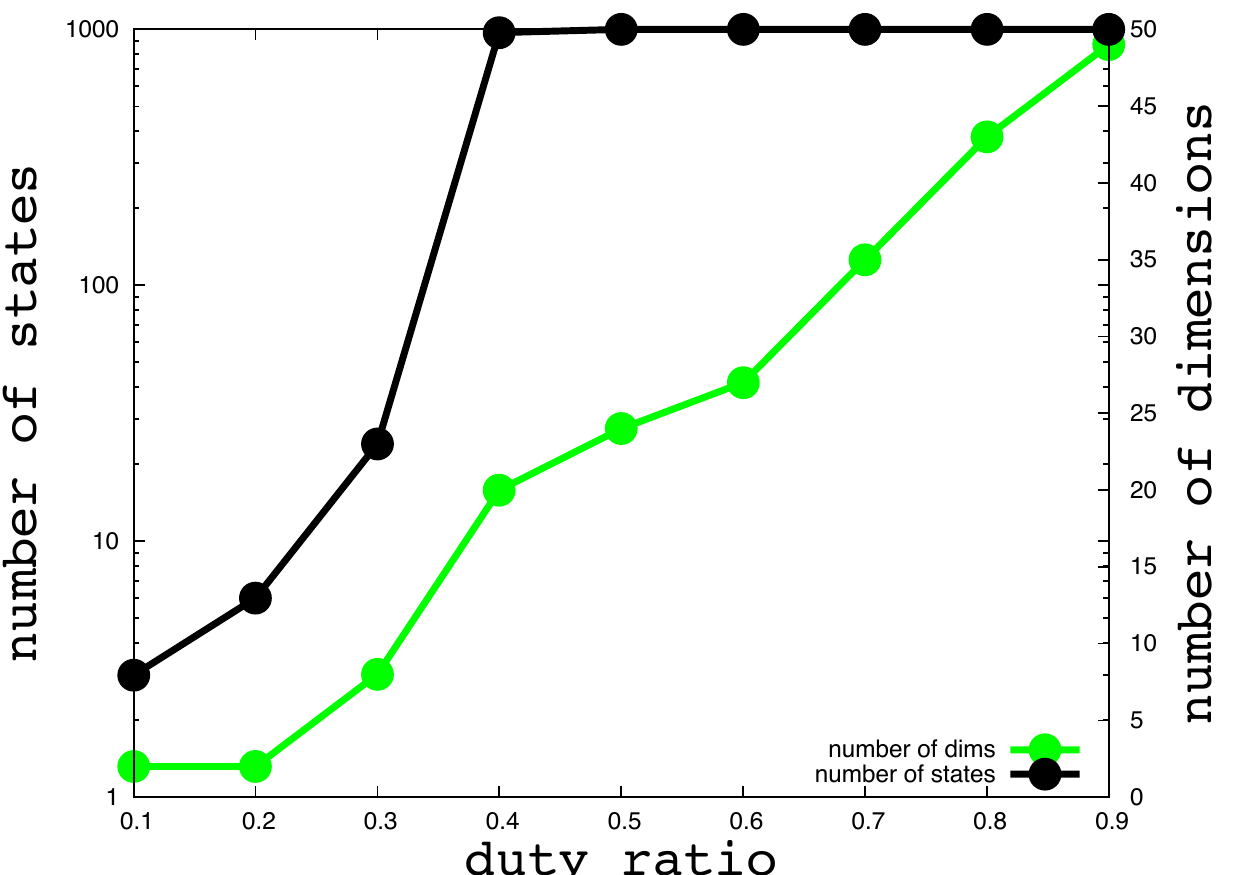}}
\subfigure[Flow = 5]{
\includegraphics[width=0.48\textwidth]{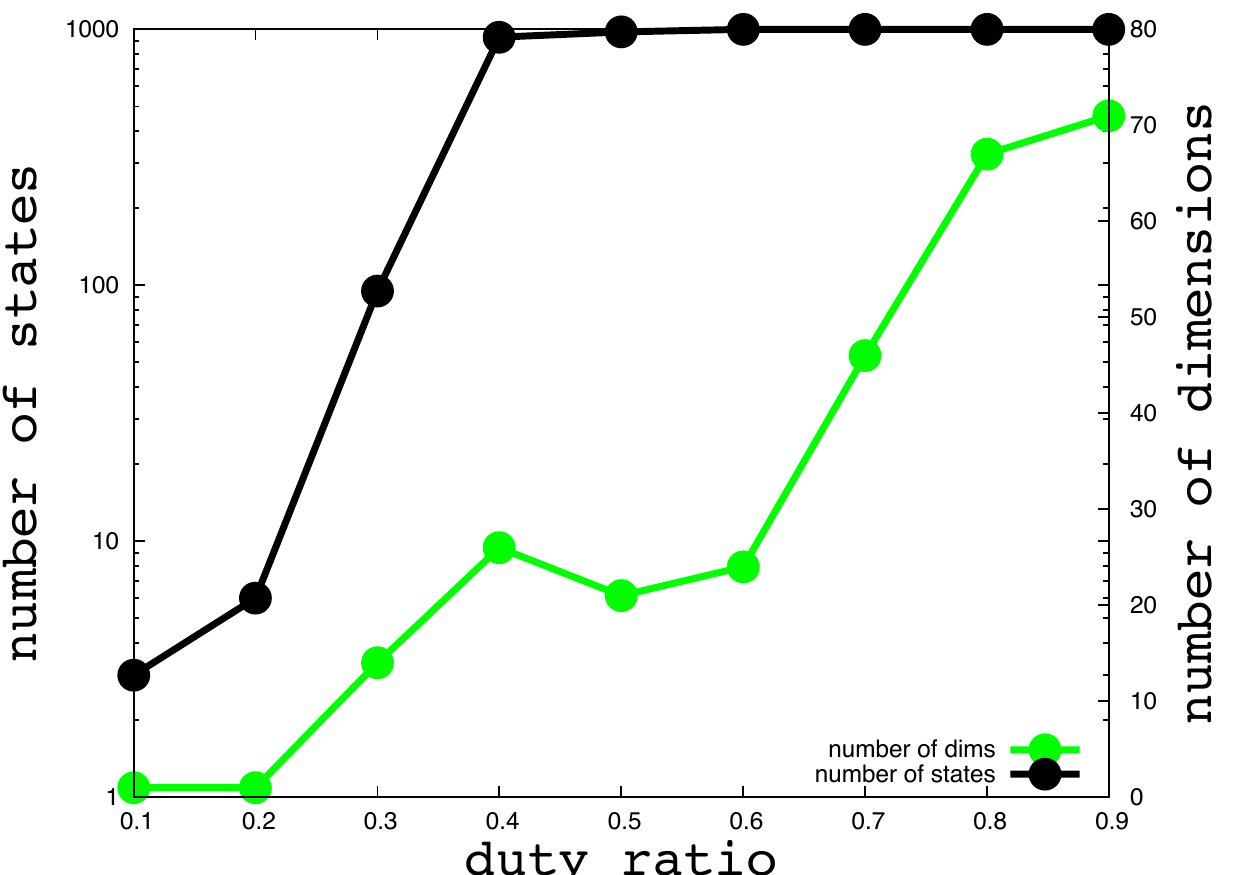}}
\subfigure[Flow = 15]{
\includegraphics[width=0.48\textwidth]{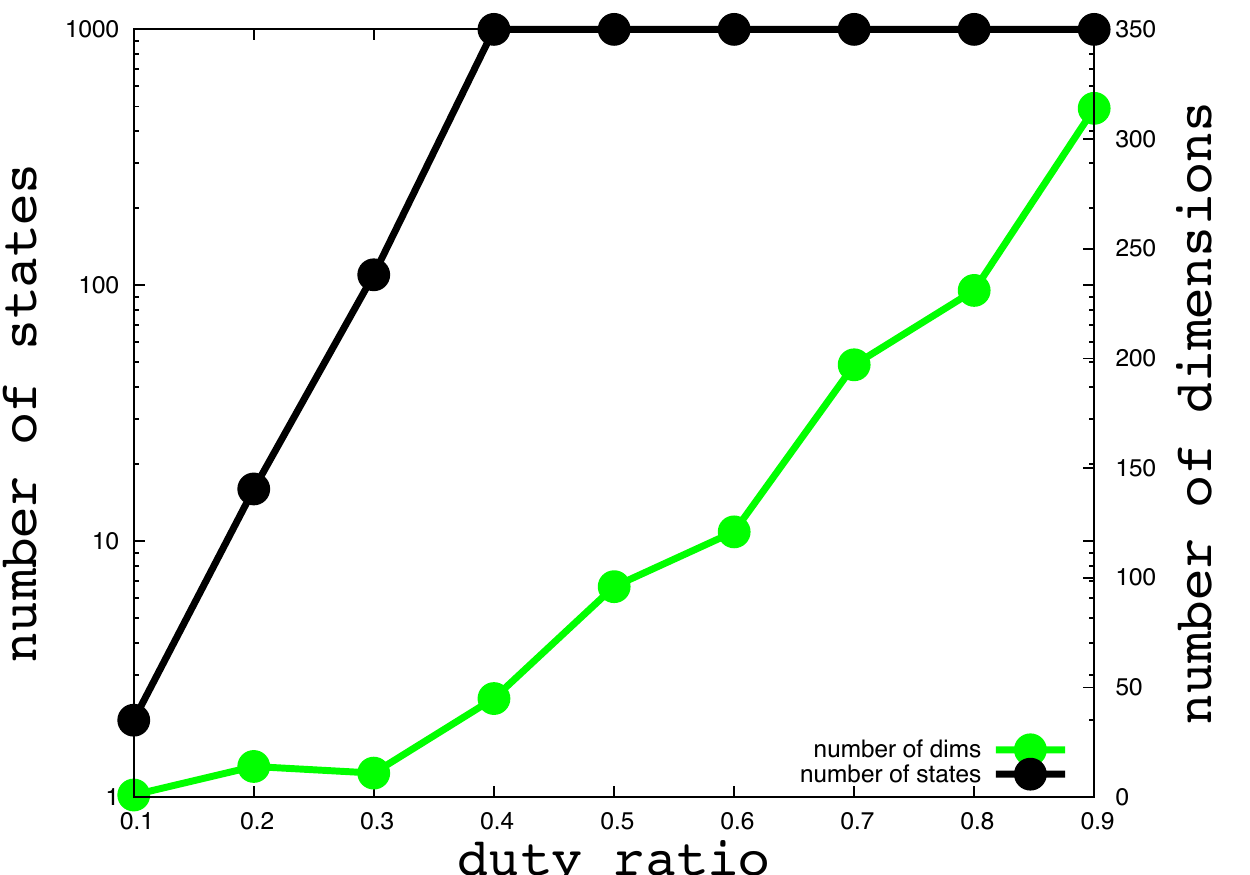}}
\subfigure[Flow = 20]{
\includegraphics[width=0.48\textwidth]{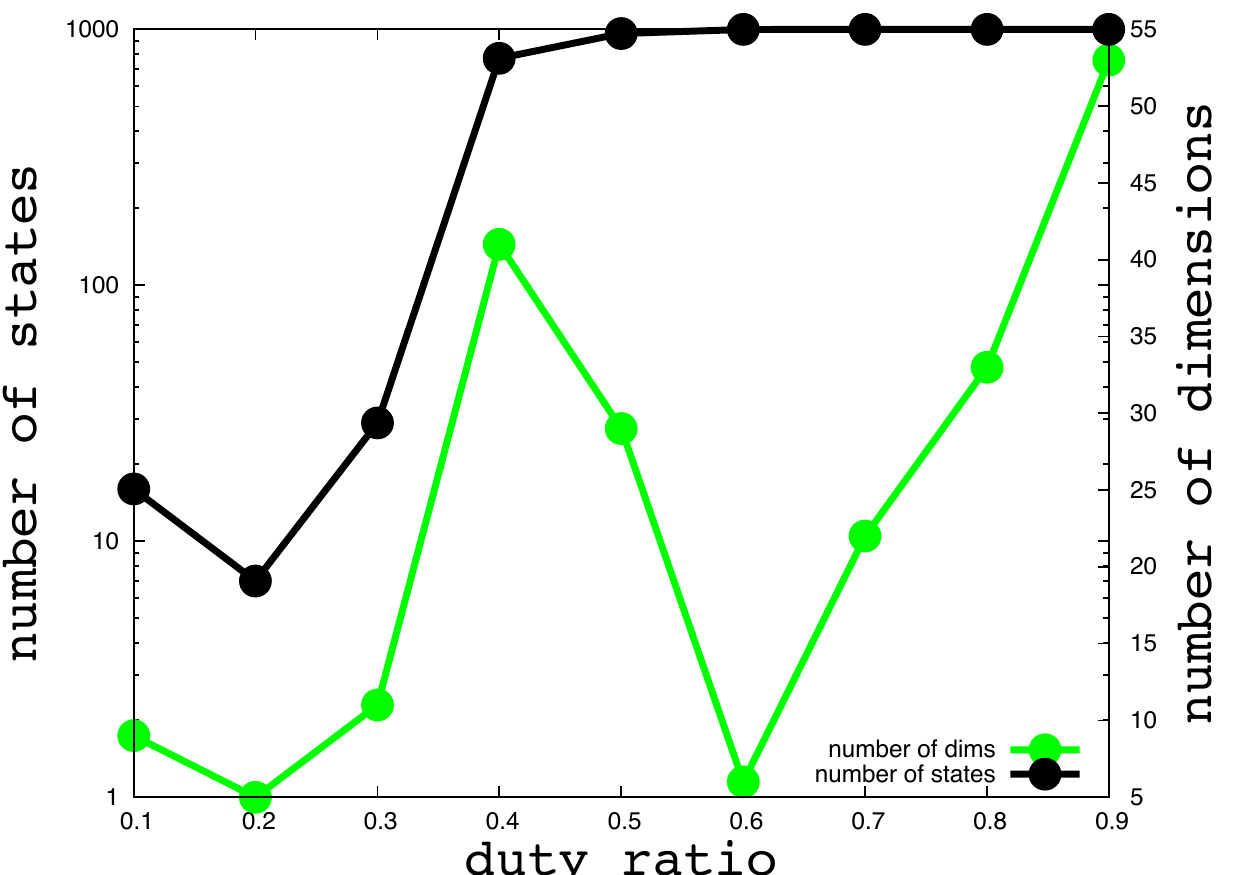}}
\caption{{\bf Change in the number of states (in black) and the number of dimensions (in green) required to exceed the contribution rate of 99\% for flows {\tt 0}, {\tt 5}, {\tt 15}, and {\tt 20}.}
\label{fig:num_states_dims}}
\end{figure}

\begin{figure}[!ht]
\centering
\includegraphics[width=0.45\textwidth]{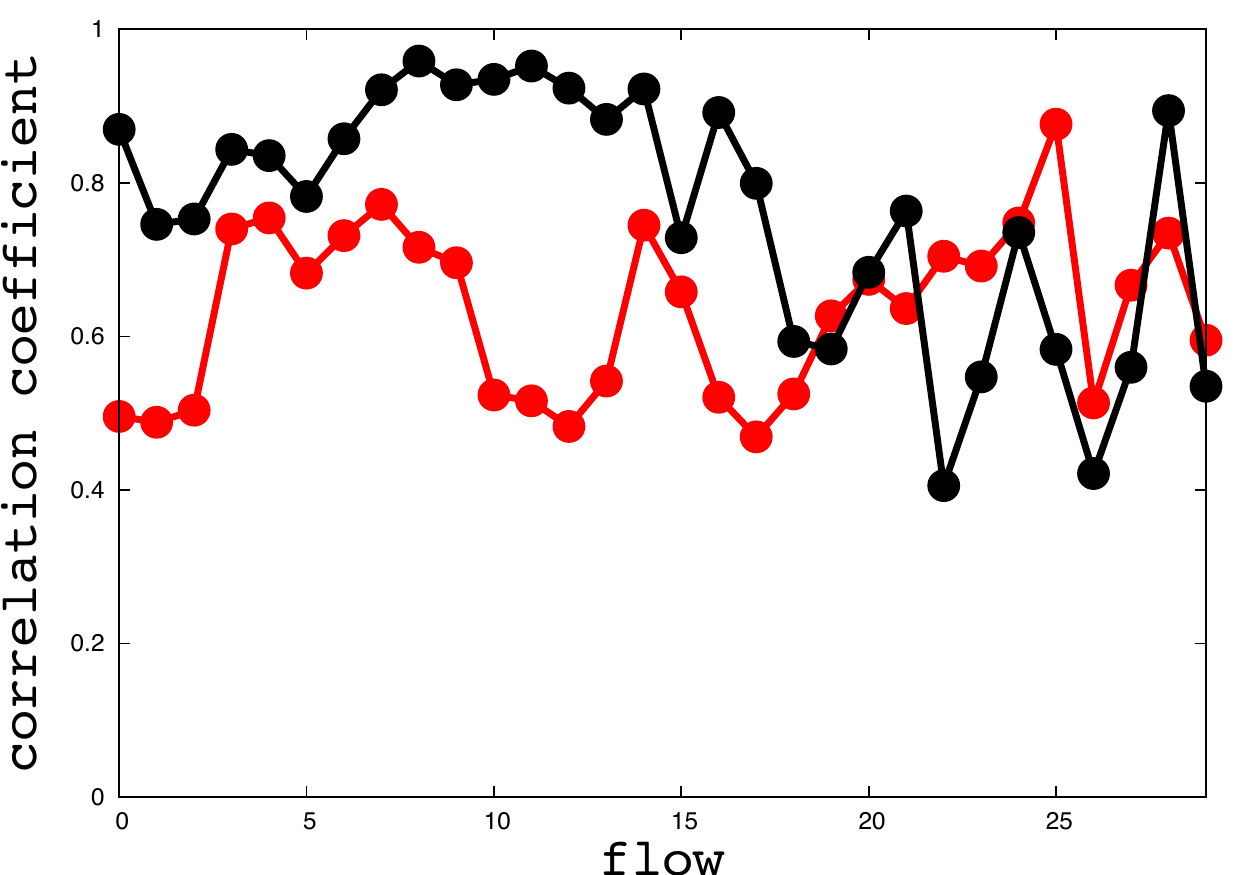}
\caption{{\bf Correlation coefficients of 30 flows.} The correlation coefficients between the number of states and the number of drops (in red) and between the number of states and the number of dimensions required to exceed 99\% of the contribution rates in the eigenvalues (in black) were computed.
\label{fig:corr_coef}}
\end{figure}

Using the transition diagrams mentioned above, we plot the changes in the number of states and the number of packet drops with respect to the duty ratio in order to find the determining factors of state stability. Figure~\ref{fig:num_states_drops} shows excerpts of the plots of flows {\tt 0}, {\tt 5}, {\tt 15}, and {\tt 20}. As the red line in Figure~\ref{fig:num_states_drops} indicates, the number of drops is positively correlated with the number of states because, as the frequency of drops increases, the cwnd dynamics become more temporally as well as spatially diverse among nodes. This results in the generation of different cwnd time series dynamics. This trend is captured in the number of dimensions of the eigenvalues required to exceed the contribution rate of 99\% of all the principal components. Figure~\ref{fig:num_states_dims} shows the changes in the numbers of dimensions and states with respect to the duty ratio. As the input ratio increases, the number of dimensions required to reconstruct the original time series increases, leading to an increase in the number of system dimensions. To validate this assertion for all 30 flows, we compute the correlation coefficient between the numbers of states and the number of drops, as well as between the number of states and the number of dimensions required to exceed 99\% of the contribution rate. According to the results, all 30 flows have positive correlations, as shown in Figure~\ref{fig:corr_coef}.

\begin{figure}[!ht]
\centering
\subfigure[Flow = 0]{
\includegraphics[width=0.48\textwidth]{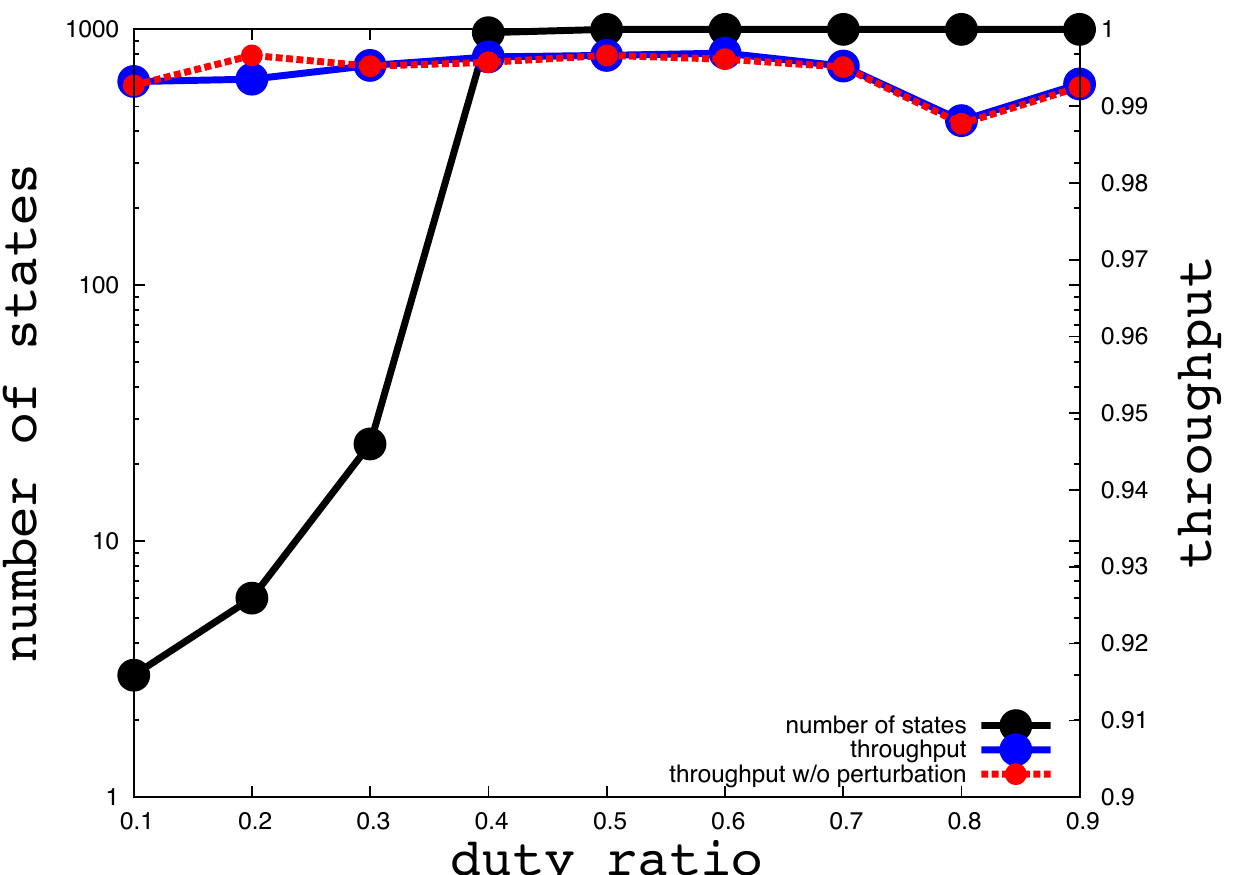}}
\subfigure[Flow = 5]{
\includegraphics[width=0.48\textwidth]{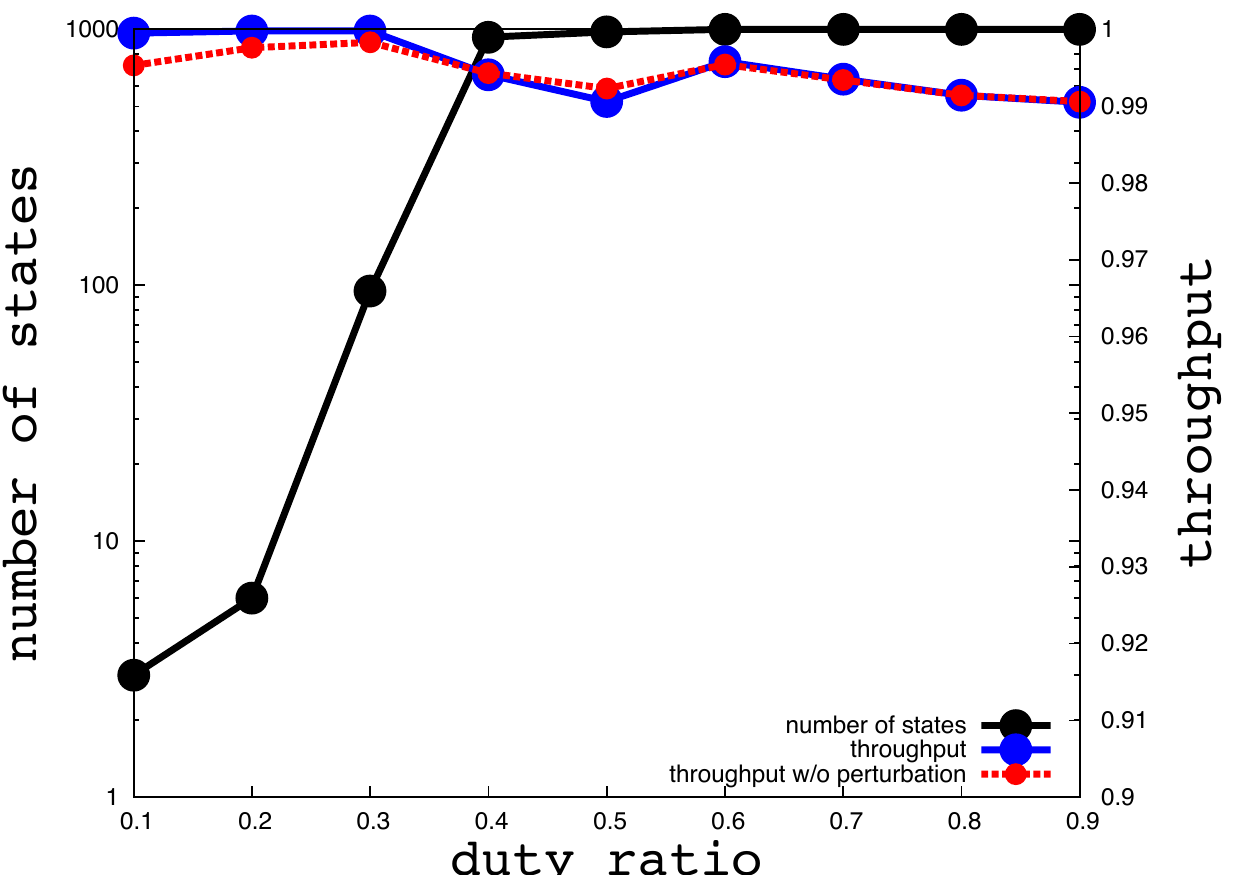}}
\subfigure[Flow = 15]{
\includegraphics[width=0.48\textwidth]{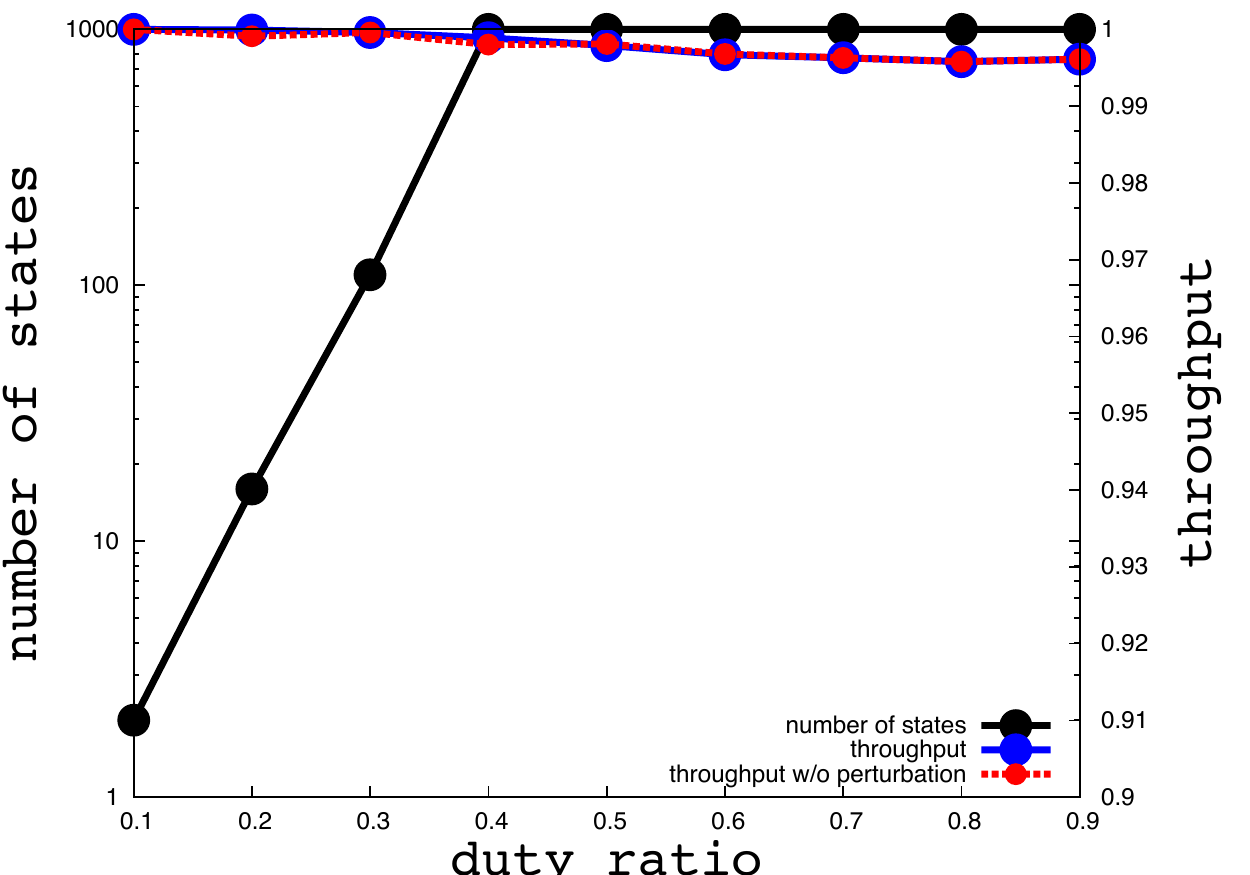}}
\subfigure[Flow = 20]{
\includegraphics[width=0.48\textwidth]{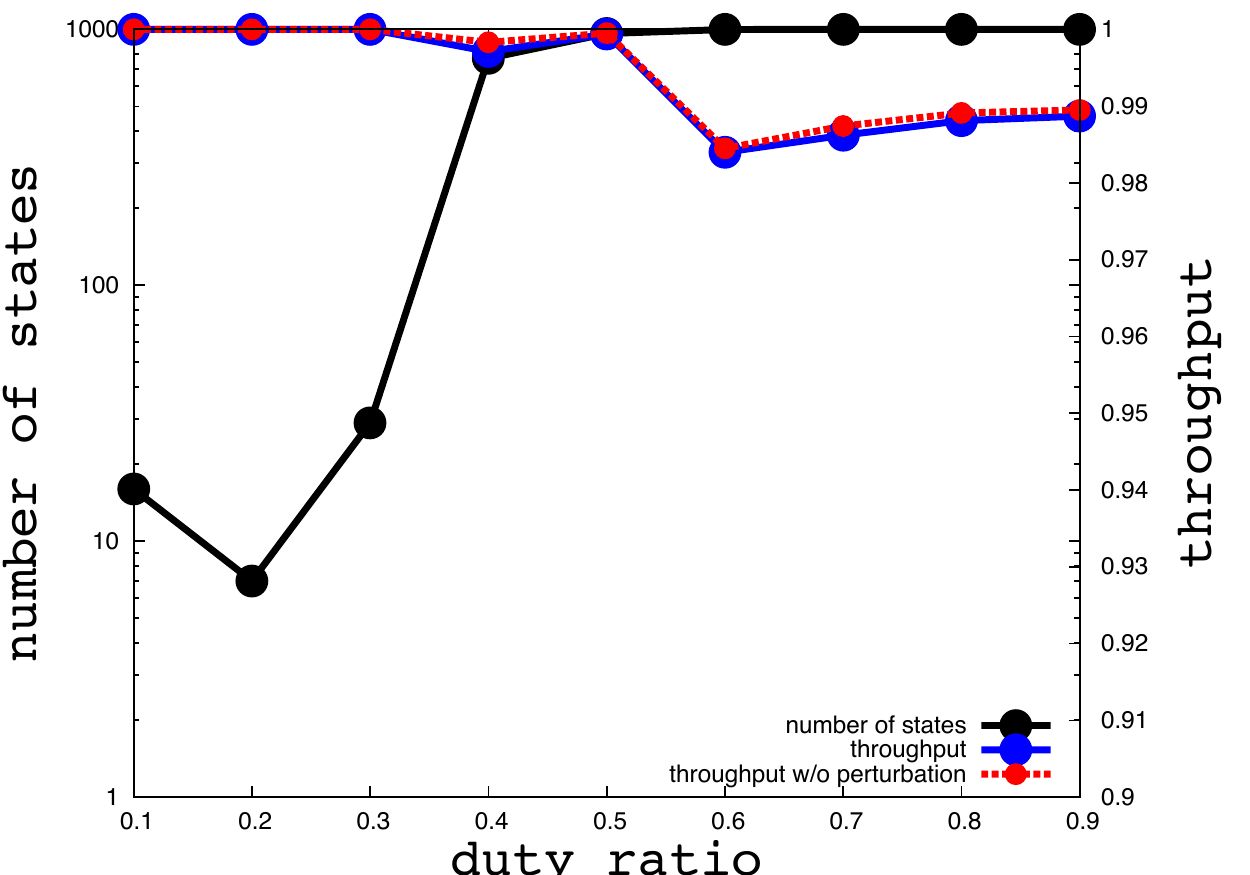}}
\caption{{\bf Change in the number of states (in black) and the throughput (in blue) for flows {\tt 0}, {\tt 5}, {\tt 15}, and {\tt 20}.}
\label{fig:num_states_throughput}}
\end{figure}

At some point (usually at around $x = 0.40$), the number of states is saturated and reaches the maximum number of states ($ = 1,000$). This change corresponds to the cwnd bifurcation diagram, which shows a state transition from a fixed or periodic point to a turbulent state (or to a large, chaotic state, since it is a deterministic system; however, as we see in Figure~\ref{fig:num_states_dims}, the dimensionality becomes too large to confirm this theory). The number of states increases suddenly when the cwnd dynamics become chaotic. When system inputs are small, the system generates fewer states and fewer transitions among the generated states. As the number of inputs increases, the system generates a greater number of states. However, interestingly, regardless of the number of states, throughput is maintained at a high rate (between 98\% and 100\%).  
The throughput is plotted (in blue) in Figure~\ref{fig:num_states_throughput} with respect to the input duty ratios of flows {\tt 0}, {\tt 5}, {\tt 15}, and {\tt 20}. In the same figure, the throughput without perturbation is also plotted (in red). There is almost no difference between the throughputs with and without perturbations. This shows the robustness of the PSN system. Note that the number of drops increases as the duty ratio increases, as plotted in Figure~\ref{fig:num_states_drops}; yet, the throughput is kept high. This is because the total number of transmitted packets increases with the duty ratio and because the ratio of drops is kept to a few percent. One may wonder what would happen  in terms of cwnd dynamics if we set up a condition with a high drop rate. Based on our analysis in this paper, it would first go to a limit cycle; cwnd would increase to a certain point but quickly decrease due to a drop, then repeat the process. Then, it would go to a chaotic state, as packet transmission timings would be gradually shifted among nodes. Eventually, certain nodes would occasionally reach a high cwnd sizes.

We can derive an important observation from these results. When the system is not congested and has a small amount of input, it creates only a few states and reverts to existing states. However, as the system's congestion increases, the system begins to generate a large number of states. Beyond a critical duty ratio, the cwnd dynamic has no more attracting states. In other words, beyond a critical point, the notion of an attractor is not applicable; instead, a process of iterating many transient states is responsible for bringing stable and better throughput on average.

\section{Discussion and Conclusions}
The PSN system described herein receives external input data, while relaying the data inside a network. The term \textit{attractor} is usually defined as the final state of development in a dynamical system; thus, an attractor does not jump to another attractor. If an attractor jumps among attractors, it is in a \textit{metastable} state or is a quasi-attractoring state. Such an unstable system nature (i.e., the intrinsic mechanism that causes this jump among metastable states), is a unique characteristic of high-dimensional dynamical systems called \textit{chaotic itinerancy}~\citep{ikegami07,kaneko,tsuda,ikeda1989}. Our work can be interpreted as proposing this chaotic itinerancy, or a network of quasi-attracting states  (i.e., attracting states from which a system eventually escapes, so that they are not attractors in a strict sense, but, rather, states in which a system can stay transitionally), as a novel way of describing PSN adaptability. In a PSN, a large number of quasi-attracting states are loosely connected as spontaneous transition processes. The loosely connected network substitutes for homeostasis, allowing for stable control even when the system is perturbed from the outside; that is, the system's behavior sustains high throughput with or without perturbation.  

Owing to Ashby and, more recently, Dyke and Weaver's works~\citep{dyke2013}, the Law of Requisite Variety has been highlighted~\citep{ashby_1958}. According to this law, a system is required to increase its internal variety in order to protect itself against perturbations. Concretely, Ashby's homeostasis model searches for an adequate attractor among many states in response to a perturbation. Similarly, Dyke and Weaver's Daisyworld model shows that the number of (fixed point) attractors increases when the number of environmental factors increases. As the result, a system can escape from an attractor smoothly to go to another one with better performance. This provides an example of Ashby's Law of Requisite Variety. 
We also show that a throughput optimized at a critical point does not capture the true picture of the PSN. Instead, increases in the number of quasi-attracting states and in the itinerant among them can account for optimal throughput over a wide range of input values. This behavior is supported by the Law of Requisite Variety. In other words, our study shows that the PSN has a property similar to that of Dyke and Weaver's Daisyworld model, thus providing another example of Ashby's law. Moreover, it shows that the PSN invalidates an attracting state in a large input region.

One may argue that the topology of the Internet is far more complex than the simulation setup in this study because nodes and edges in the Internet are added or removed constantly, thereby causing unexpected packet flows. Nevertheless, our PSN experiment, conducted with a simple network topology, has important implications because its discussion of PSN attractors is also applicable to more complex network cases. As for future works, an investigation of the dynamics of PSNs with growing network topologies and with the automatic route-searching dynamics of PSN would be worthwhile. Various packet data flows corresponding to different applications would represent the practical interests (e.g., short text messages often used in social network services or the streaming of movies). One could also study real PSN data by gathering packets flowing in the actual Internet and comparing the results with the results presented here.  

Our findings could be useful in designing a protocol for a more recent attempt at packet control by a programmable protocol, such as OpenFlow~\citep{OpenFlow}. In the previous studies, we examined the information dynamics of social network services in order to draw an analogy between brain dynamics and the real Internet~\citep{oka2013}. If we can further develop the present studies of Ashby's law with regard to the real-world Internet, we will make a truly biologically- (brain science) inspired model and research summary of the Internet (i.e., the RYF property of the Internet).

\section*{Acknowledgements}
This work was supported by the Japan Society for the Promotion of
Science Grant-in-Aid for Young Scientists (B) (\#25730184 ``Burst
Analysis of Twitter time series based on RT diffusion and its
application to Web services'' and partially by \#24700031 ``Wide-area
distributed archival storage with replica allocation mechanism based
on throughput prediction'') and partially by Grant-in-Aid for
Scientific Research on Innovative Areas (\#24120704 ``The study on the
neural dynamics for understanding communication in terms of complex
hetero systems'') as well as Grant-in-Aid for RIKEN ``Promotion of the
Integrated Medical Science'' (2012-13).  The funders had no role in
study design, data collection and analysis, decision to publish, or
preparation of the manuscript.

\bibliographystyle{harvard}
\bibliography{biblio}

\end{document}